\documentclass[a4paper,fleqn,usenatbib]{mnras}

\usepackage[T1]{fontenc}
\usepackage{ae,aecompl}

\usepackage{graphicx}	
\usepackage{amsmath}	
\usepackage{amssymb}	
\graphicspath{{./plots/}}

\newcommand{\R}{$f_\mathrm{em}/f_\mathrm{abs}$}

\newcommand{\Ha}{H$\alpha$}

\newcommand{\ek}{$E_\mathrm{k}$}

\newcommand{\lp}{$L_\mathrm{p}$}
\newcommand{\trise}{$t_{-1/2}$}
\newcommand{\tdecay}{$t_{+1/2}$}
\newcommand{\eom}{$E_\mathrm{k}/M_\mathrm{ej}$}

\newcommand{\tmax}{$t_\mathrm{max}$}

\newcommand{\mni}{$M_\mathrm{Ni}$}

\title[Classification of SE-SNe]{A physically motivated classification of stripped-envelope supernovae}

\author[S. J. Prentice]{
S. J. Prentice,$^{1}$\thanks{E-mail: S.J.Prentice@2014.ljmu.ac.uk}
P. A. Mazzali,$^{1,2}$
\\
$^{1}$Astrophysics Research Institute, Liverpool John Moores University, IC2, Liverpool Science Park, 146 Brownlow Hill, \\  Liverpool L3 5RF, UK\\
$^{2}$Max-Planck-Institut f{\"u}r Astrophysik, Karl-Schwarzschild-Str. 1, D-85748 Garching, Germany\\
}

\date{Accepted XXX. Received YYY; in original form ZZZ}

\pubyear{2016}

\begin{document}
\label{firstpage}
\pagerange{\pageref{firstpage}--\pageref{lastpage}}
\maketitle

\begin{abstract}
The classification of stripped-envelope supernovae (SE-SNe) is revisited using modern data-sets. Spectra are analysed using an empirical method to ``blindly'' categorise SNe according to spectral feature strength and appearance. This method makes a clear distinction between SNe that are He-rich (IIb/Ib) and He-poor (Ic) and further analysis is performed on each subgroup. For He-rich SNe the presence of H becomes the focus. The strength, velocity, and ratio between absorption and emission of H$\alpha$ is measured, along with additional analysis of \ion{He}{I} lines, in order to categorise the SNe. The He-poor SNe are ordered according to the number of absorption features $N$ present in the spectra, which is a measure of the degree of line blending. The kinetic energy per unit mass $E_\mathrm{k}/M_\mathrm{ej}$ is strongly affected by mass at high velocity and such situations principally occur when the outer density profile of the ejecta is shallow, leading to the blending of lines. 
Using the results, the existing SE-SN taxonomic scheme is adapted. He-rich SNe are split into four groups, IIb, IIb(I), Ib(II), and Ib, which represent H-rich to H-poor SNe.
The SNe Ic category of broad-lined Ic (Ic-BL) is abandoned in favour of quantifying the line blending via $\left<N\right>$ before peak. 
To better reflect the physical parameters of the explosions, the velocity of \ion{Si}{II} at peak and the half-luminosity decay time $t_{+1/2}$ are included to give SNe Ic a designation of Ic-$\left<N\right>\left(v_\mathrm{p,SiII}/t_{+1/2}\right)$.
\end{abstract}



\section{Introduction} \label{sec:1}
Stripped-envelope supernovae (SE-SNe) \citep{Clocchiatti1997} are a subset of core-collapse SNe where the progenitor star has lost much of its outer layers of hydrogen and in some cases helium. Classically, these SNe have been classified as Type IIb for those SNe that are H-rich, Type Ib for H-poor/He-rich, and Type Ic for H-poor/He-poor as given by the features in their optical spectra \citep[See][]{Filippenko1997,Matheson2001}. 

Type Ib SNe were first identified in the 1980s with the discovery of SNe 1983I, 1983N, 1983V, 1984L, and 1985F. They were found to have a light curve evolution similar to that of a Type Ia SN but without the post-peak bump in the redder bands \cite{Elias1985}. They also showed prominent He lines in the spectra \citep[e.g.,][]{Wheeler1985,Harkness1987,Wheeler1986,Filippenko1986,Gaskell1986}, a situation irreconcilable with a CO white dwarf progenitor. The He lines could only come from a stellar envelope, indicating that the SN were of core-collapse origin. Yet, they lacked the prominent H lines in the spectrum and the long diffusion timescales that characterise Type II light curves, meaning they lacked the high mass H envelope of ``typical'' Type II SNe and on this basis were labelled as Type Ib SN. 

Evidence of another type of stripped envelope SN came with the discovery of SN 1993J \citep{Nomoto1993,Filippenko1993} which initially showed prominent Balmer series lines with clear P-Cygni profiles in its spectra but evolved to look spectroscopically similar to Type Ib SNe. The light curve of SN 1993J showed an initial rapid declining phase from a luminosity indicative of the cooling post-shock breakout from the stellar envelope \citep{Woosley1994}. It then proceeded to rise again over the following nine days, a behaviour typical of Type I SNe which are powered by $^{56}$Ni. The spectral and photometric evolution and properties of a SN with these properties had previously been theorised as a ``Type IIb'' event \citep{Ensman1987}. However, the relationship between SNe Ib and SNe IIb is not clear and it has been suspected that some Type Ib SNe \citep[e.g., SN 2005bf][]{Tominaga2005} show evidence of hydrogen in their optical spectra hinting that there may be at least some overlap between the two populations.

Type Ic SNe were identified as having somewhat similar spectra to Type Ia SNe but without the deep \ion{Si}{II} 6347 \AA\ line and were somewhat dimmer than normal SNe Ia.  SN 1994I \citep{Nomoto1994,Iwamoto1994,Filippenko1995} was the first well-observed object of this class and has long been considered an example of a ``normal'' SN Ic. 
The discovery of SNe 1997dq \citep{Mazzali2004} and 1997ef \citep{Mazzali2000,Iwamoto2000} introduced a new category of ``broad-lined'' Ic (Ic-BL), which showed broad, shallow absorption lines and high velocity ejecta. Shortly after these SNe were discovered came the first example of a SN associated with a gamma-ray burst (GRB), SN 1998bw/GRB980425 \citep{Galama1998,Iwamoto1998,Patat2001}. Gamma-ray burst supernovae (GRB-SNe) fall exclusively into the broad-lined category. 
The spectra of broad-lined SNe Ic are characterised by an almost featureless red spectrum with significant line blending and blanketing that belies the temperature of the actual photosphere due to the reprocessing of photons through scattering processes. Line broadening is a product of high velocity material with sufficient mass to provide opacity over a large velocity range, a situation that results in high kinetic energy ejecta ($E_\mathrm{k} > 10^{52}$ erg). To date at least a dozen 1998bw-like Ic-BL SNe have been discovered (e.g., SN 2002ap \citep{Mazzali2002,Foley2003}, SN 2003dh/GRB030329 \citep{Stanek2003,Mazzali2003,Matheson2003}, SN 2006aj/XRF060218 \citep{Pian2006,Mazzali2006,Modjaz2006,Sollerman2006}, SN 2010bh/GRB100316D \citep{Cano2011,Bufano2012}, SN 2013cq/GRB130427A \citep{Xu2013,Melandri2014}), of which several have been associated with either a gamma-ray burst or an X-ray flash (XRF). In addition it is not possible to fully discount the possibility of a GRB association for SN 1997ef or SN 2002ap. The Ic-BL moniker has also been attributed to SNe Ic with lines broader than that of SN 1994I resulting in ambiguity between Ic SNe that genuinely share spectroscopic similarity with SN 1998bw and those which are moderately energetic.

The degree of stripping of a SN has important implications for its progenitor and the system that hosted it. 
It has been suggested for some time that the rates of SE-SNe are irreconcilable with the IMF if they are the products of single star (Wolf-Rayet) systems.
This is because for a star to lose most of its H envelope due to winds it must be more than at least 20 $M_\odot$ with solar metallicity or considerably more massive if $Z$ is lower \citep[See, for example][]{Hirschi2004,Crowther2007,Smith2014}. 
The solution to this problem is likely found in interacting binaries whereby a common envelope phase robs the progenitor star of much of its outer envelope \citep[e.g.,][]{Nomoto1994,Eldridge2013,Yoon2015}. This process allows for SE-SNe to occur from lower mass progenitors.

This study is motivated by the need to reassess the present taxonomy with more modern data sets as the classification system presently used developed before there was a sufficiently large number of SNe for each sub-class. In principle we aim to address several issues. 
The first is how SNe IIb relate to SNe Ib; it can be taken that there are extreme examples of these SNe but in an era with considerably more spectral information, do the labels ``IIb'' and ``Ib'' reflect the actual properties of the SNe? 
If SNe IIb are H-rich and SNe Ib H-poor is there a way to reflect this in the classification scheme? SNe Ic are considered in the context of line broadening with the aim of answering the question ``Just how broad are broad lines?'' by quantifying the degree of breadth of the absorption lines in the SNe Ic spectra. It is anticipated that this will provide clarity in a field where the label `Ic-BL' is arbitrary. 
Categorisation of SNe using spectral templates, for instance via the Supernova identification tool SNID \citep{Blondin2007}, is useful but spectral similarity does not equate to similar explosion properties. For example, several SNe have shown spectral similarity to the peculiar Type Ib SN 2005bf (e.g., SN 2015ah and SN 2016frp) but as yet none have shown the peculiar spectroscopic and photometric evolution of SN 2005bf. Thus, the process of comparing spectra with reference spectra should be the beginning of the classification procedure which may last some weeks as more data is collected. Consequently, the intention is to move taxonomic schemes away from spectral template similarity which impose a bias upon object categorisation and data collection. A useful classification system should be simple and yet provide information about the properties of the object (for instance, in the way that stellar classification does). To this end we will seek to adapt the existing frame-work to be more representative of SE-SNe and their characteristics. 

This study utilises mostly publicly available data. In Section~\ref{sec:2} the initial method for analysing the SNe spectra is described. In Section ~\ref{sec:3} the presence of H in SNe IIb and Ib is analysed and quantified using an analytical method based upon the properties of the H$\alpha$ line, and in Section~\ref{sec:4} the findings of this study are presented. Section~\ref{sec:5} details the analytical method and results for SNe Ic, while Section~\ref{sec:6} presents the results and introduces a method for classifying SNe Ic based upon the number of absorption features visible in the optical spectra. In Section~\ref{sec:8} a final adjustment made to the taxonomy of SNe Ic, based partly upon work presented in Appendix~\ref{sec:7}. In Section~\ref{sec:9} the results for both He-rich and He-poor SNe are discussed and finally, the study is concluded in Section~\ref{sec:10}.

\section{Method - Visual inspection} \label{sec:2}
The SNe analysed here are listed in Table~\ref{tab:database} and were mostly selected from the SE-SNe database of \cite{Prentice2016} so as to provide comparative light curve properties, in particular the epoch of each spectrum with reference to bolometric maximum. 
We increased this sample by also including Ic-BL SN 1997ef \citep{Mazzali2000,Iwamoto2000} and Type Ic SN 2004dn \citep{Drout2011,LAW2012} as the former is an interesting object spectroscopically and the latter had photometry that was not previously available. 
We also include preliminary analysis on Type Ic SNe 2012ej, 2016P, 2016coi\footnote{aka ASASSN-16fp}, and 2016iae (Prentice et al. in preparation).  
The public spectra for each SN were downloaded from WISeRep\footnote{http://wiserep.weizmann.ac.il/} \citep{Yaron2012} and supplementary photometry from the Open Supernova Catalog (OSC)\footnote{https://sne.space/} \citep{Guillochon2016}.

Each spectrum was shifted to the rest-frame wavelength as given in Table~\ref{tab:database} however the spectra were not corrected for any kind of extinction, for two reasons. The first is that, even though Galactic extinction is known, host extinction is usually unknown and as host extinction normally dominates Galactic extinction then the bulk of the total extinction is unknown. 
Secondly, the analysis performed here relies on relative flux rather than absolute flux and is derived over $\sim1000$ \AA\ regions which are small enough to mitigate the differences associated with extinction differential over wavelength. 
However, for aesthetic reasons extinction corrections were applied to plotted spectra if $E(B-V)>0.1$ mag.

The next step was to visually inspect the spectra in our sample in bins of 5 days, from $-12.5$ to $+12.5$ days from bolometric maximum. If there were multiple spectra in each bin we chose the spectrum closest in time to the mid-point however, we also considered S/N and would prefer a spectrum if it was significantly better than the others. To mitigate the possibility of bias in the grouping, the spectra were not labelled with either name or classification. 

We grouped the spectra along two axes; hydrogen line strength, and broadness of the lines. As would be expected this created an immediate separation between SNe of Type Ic and SNe of Type Ib/IIb with no migration of SNe between the two groups. We then followed two separate analytical pathways which we describe in Section~3 for the SNe Ib/IIb (He-rich) and Section~5 for the SNe Ic (He-poor).

\section{He-rich SNe: The issue of hydrogen} \label{sec:3}
The presence of hydrogen in SE-SNe has previously been discussed in the literature with \cite{Elmhamdi2006} and \cite{Deng2000} using SYNOW to identify H$\alpha$ in various SNe Ib while \cite{Branch2006} and \cite{Parrent2016} find that SNe Ic as well as SNe Ib may have some remaining H. \cite{Tominaga2005} found signs of hydrogen in the peculiar Type Ib SN 2005bf while \cite{Fremling2016} presented the possibility of weak H in the optical and infrared spectra of SN Ib iPTF13bvn. \cite{Folatelli2014} suggested using used the H velocity profile as a way of separating SNe IIb and H-rich SNe Ib into different groups. \cite{Hachinger2012} used non-local thermodynamic equilibrium (NLTE) models to determine that less than 0.1$M_\odot$ of H can be present in SNe Ib before H lines begin to appear, but they speculated that the H$\alpha$ line may be a H$\alpha$/\ion{Si}{II} blend. 

\begin{figure}
	\centering
	\includegraphics[scale=0.4]{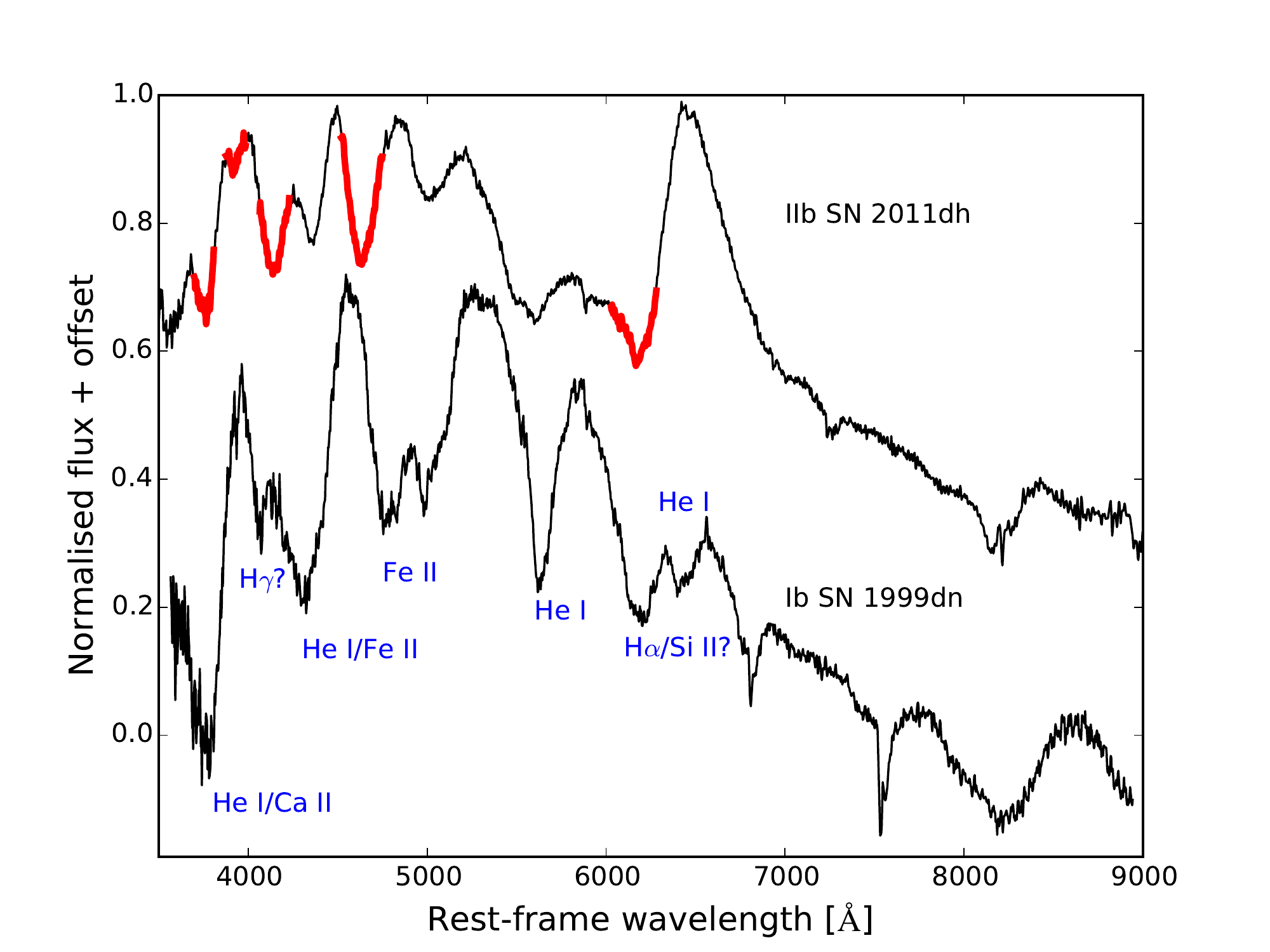}
	\caption{A comparison of the pre-peak spectra of Type IIb SN 2011dh and Type Ib SN 1999dn. Highlighted in red on the SN 2011dh spectrum are the Balmer series of absorption lines associated with H$\alpha-\epsilon$ which indicate a H rich outer ejecta. Comparatively, in SN 1999dn, aside from a feature in a similar position to H$\alpha$ there is no clear indication of the higher Balmer lines. }
	\label{fig:Hdemo}
\end{figure}

Figure~\ref{fig:Hdemo} shows the canonical Type IIb SN 2011dh \citep{Bersten2012,Arcavi2011,Ergon2014,Soderberg2012}; the P-Cygni profiles of the hydrogen Balmer series are prominent in the spectra. On the other hand we can also see SN Ib 1999dn \citep{Deng2000,Benetti2011} for which it is unclear if there is H present in the spectra. 
The P-Cygni profile at the position of the H$\alpha$ line ($\sim 6200$ \AA) is weak and there is no indication of the higher Balmer lines. This line could also be due to \ion{Si}{II} $\lambda$ 6347 as in SNe Ic and Ia; a discussion of this can be found in Appendix~\ref{sec:si}, however we will refer to this feature as \Ha\ throughout the discussion of He-rich SNe.

\subsection{H and He Line velocities}
The velocity evolution for \ion{H}{} and \ion{He}{} is calculated from the H$\mathrm{\alpha}$ and \ion{He}{I} 5876 \AA\ lines using the maximum depth of the absorption feature and in most cases the lines were sufficiently well defined with no evidence of blending. However, at early times and shortly after maximum the features can be blended or less pronounced. In a situation whereby an absorption feature shows two minima we would consider the later evolution of the feature in order to attribute the minima to the correct species. The line velocities for each SN were fit with a low order spline and an approximate velocity was calculated for every day relative to maximum light. This was then used to calculated the mean velocity of all the SNe in each daily bin.

\begin{figure}
	\centering
	\includegraphics[scale=0.7]{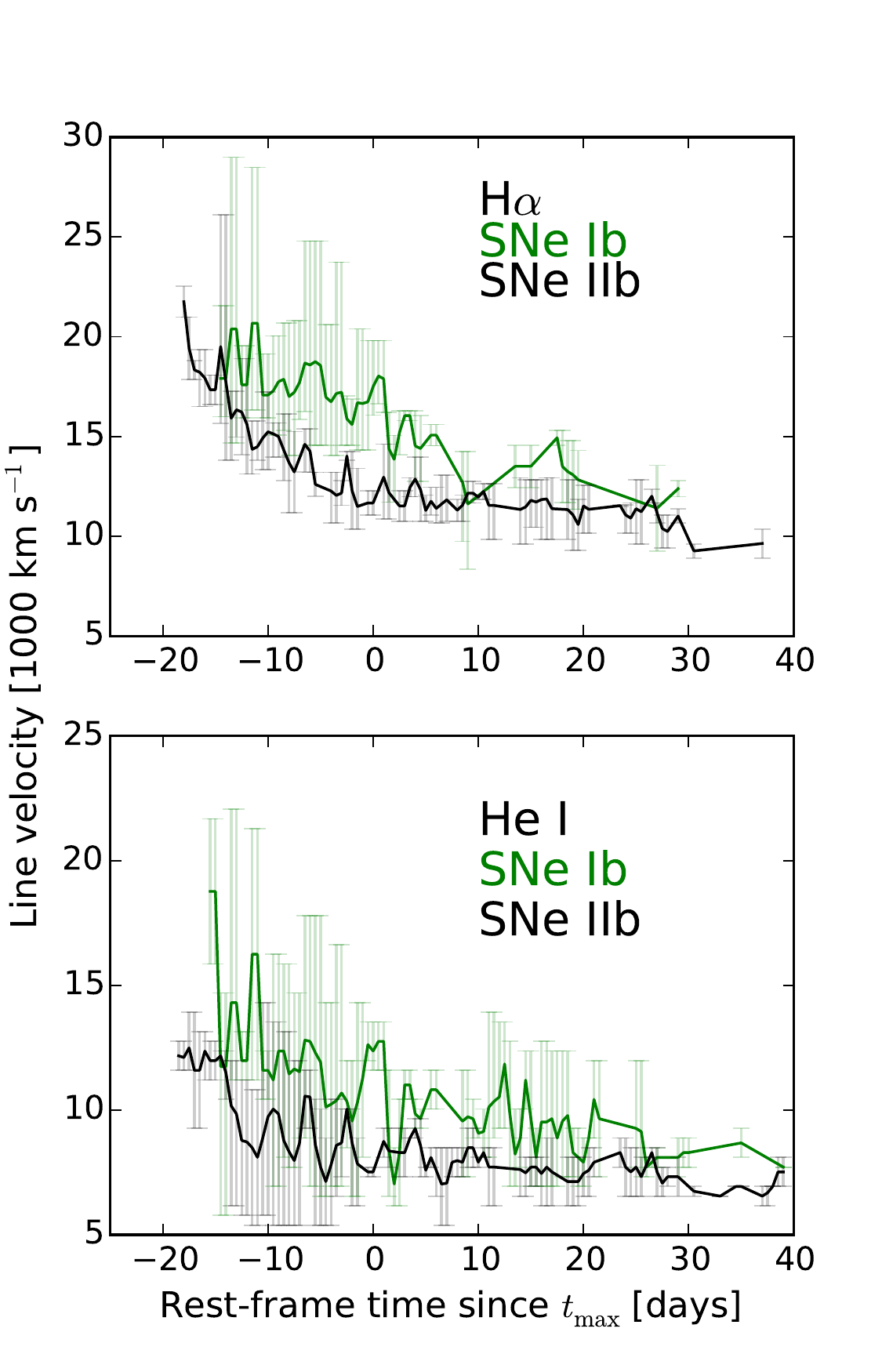}
	\caption{(Top) The mean velocity of \ion{H}{} as given by the H$\alpha$ line for SNe IIb (black) and SNe Ib (green). The error bars represent that range of values in each daily ``bin''.  This indicates there is a continuum of values however the means of the supernovae subclasses are distinct. (Bottom) Mean \ion{He}{} velocities as derived from measuring the \ion{He}{I} 5876 \AA\ line. These lines show far more overlap in the values between subgroups than for hydrogen. }
	\label{fig:vels}
\end{figure}

The velocity measurements are shown in Figure~\ref{fig:vels} for H$\alpha$ and \ion{He}{I} 5876. The velocities are broadly consistent with those found in \cite{Liu2016} for a slightly different data set. There is a continuum of $v_\mathrm{H}$ values but little overlap between SN sub-types. The situation for \ion{He}{I} is different as the velocities do show considerable overlap, although some SNe Ib can show higher line velocities in individual cases. The \ion{He}{I} 5876 measurement is of importance because the identification of this line is not ambiguous and will provide a useful reference point with regard to the structure of the SN ejecta.

Finally, the H$\alpha$ feature itself is transient, especially in SNe Ib, and its lifetime can be limited to $\sim10$ d past maximum light.

\subsection{H$\alpha$ and He I 5876 Equivalent width} \label{sec:ew}
\begin{figure}
	\centering
	\includegraphics[scale=0.4]{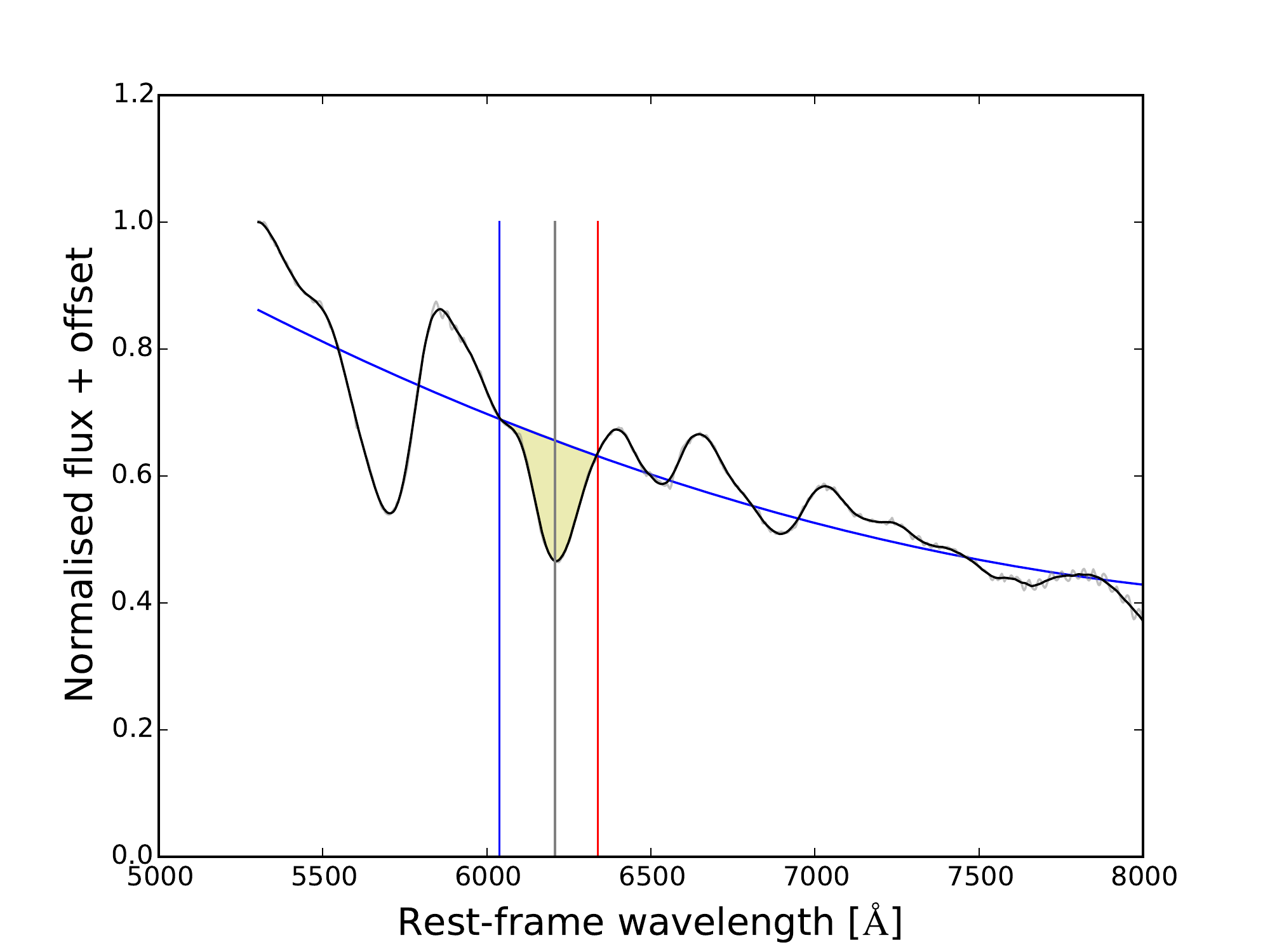}
	\caption{A demonstration of the results of the equivalent width fitting code. A pseudo-continuum, either linear or quadratic, is fit to the normalised and smoothed spectrum around the absorption feature. The yellow shaded region is then used to calculate $EW$.}
	\label{fig:EWdemo}
\end{figure}

The equivalent width $EW$ for H$\alpha$, and for comparison \ion{He}{I} 5876, is calculated in order to establish the temporal evolution of the line strength and to provide another comparative characteristic between the lines. The continuum level for the wavelength region in question was approximated by fitting the spectra with either a quadratic or linear spline in a range of a few thousand angstroms either side of the feature. Each continuum fit was inspected by eye and in most cases provided a reasonable fit to the continuum level. In the few instances where the fit was clearly inaccurate the spectrum was removed from the sample. The $EW$ is then:

\begin{equation}
EW = \int_{\lambda_a}^{\lambda_b} \left[1 - F(\lambda)/F_\mathrm{c}(\lambda)\right] d\lambda
\end{equation}
where $\lambda_a$ and $\lambda_b$ are the wavelength boundaries, $F(\lambda)$ is the flux of the spectrum at $\lambda$ and $F_\mathrm{c}(\lambda)$ is the continuum flux at $\lambda$. The largest sources of uncertainty come from the continuum fit and the boundary limits. In the former case the problem arises when there are significant emission peaks masking the underlying continuum. In the latter case one would ideally want to place limits at the edges of the absorption feature but multi-component lines require some estimation of where the boundary is. Fortunately these two uncertainties are rarely seen together, as spectra with prominent emission also tend to have well defined absorption profiles and spectra with multi-component lines tend to have less emission. A demonstration of this procedure is shown in Figure~\ref{fig:EWdemo}.

\begin{figure}
	\centering
	\includegraphics[scale=0.7]{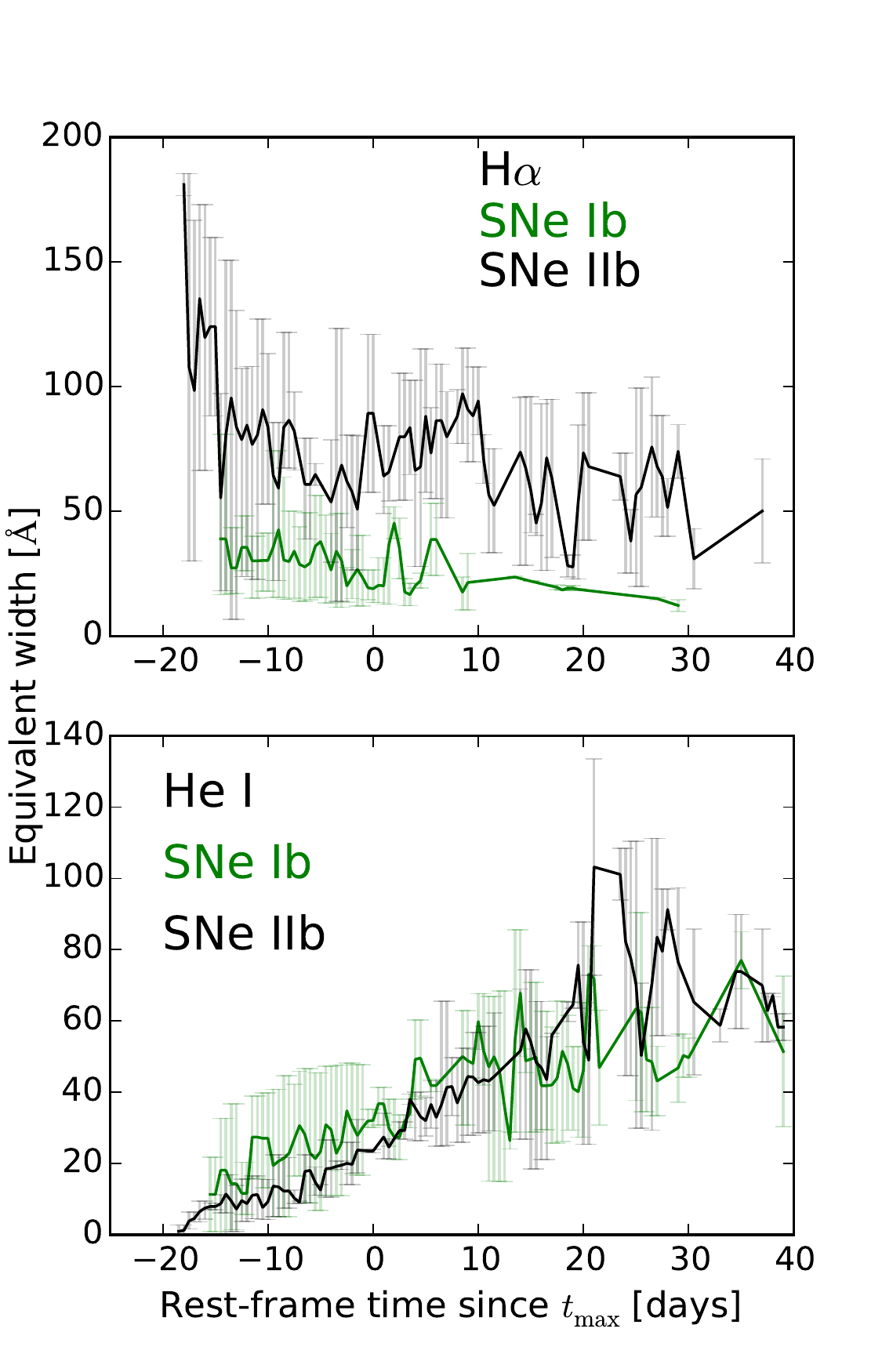}
	\caption{(Top) The mean $EW$ as a function of time for H$\alpha$, the strong H lines of the SNe IIb is prominent but there is overlap with the stronger H$\alpha$ lines of the SNe Ib sample. (Bottom) The same but for \ion{He}{I} 5876, this line grows in strength over time but is especially strong in some SNe Ib. In both plots the error bars represent the maximum range of values used to calculate the mean.}
	\label{fig:EW}
\end{figure}

The curves of the total mean $EW$ of H$\alpha$, and \ion{He}{I} 5876, for the He-rich SNe are shown in Figures~\ref{fig:EW}. Our results are consistent, in terms of relative positioning of the $EW$ curves of the SN types, with \cite{Liu2016} for their sample. Our actual values differ but this can be explained by the different methods used to process the spectra and calculate $EW$.

In Figure~\ref{fig:EW} it can be seen that $EW$ of H$\alpha$ is weak in some SNe IIb shortly after explosion. 
This is because the temperature of the ejecta at this epoch is sufficient to ionize H and so there is little neutral H above the photosphere to provide line opacity. 
As the photosphere cools we see the absorption increase in strength to around bolometric maximum. For the SNe Ib the feature is typically weaker although the line strength of some SNe Ib overlaps with that of the weaker SNe IIb.

The $EW$ of the \ion{He}{I} 5876 line (Figure~\ref{fig:EW}) is quite similar for most of the SNe with the line increasing in strength over time. This occurs because the photosphere gradually exposes the denser regions of He in addition to the increased $\gamma$-ray flux as the photosphere moves towards more $^{56}$Ni rich regions which allows non-local thermodynamic equilibrium (NLTE) excitation of the exposed He \citep{Lucy1991}.

\subsection{H$\alpha$ emission to absorption ratio \R}
As previously noted, the SNe were ordered based upon the relative strength of the H features, especially H$\alpha$, and it is apparent that there is a diversity of spectral shape within this region. In particular the relative intensity of the absorption component to the emission component varies amongst SNe and as a function of time within each SN. 

\begin{figure}
	\centering
	\includegraphics[scale=0.4]{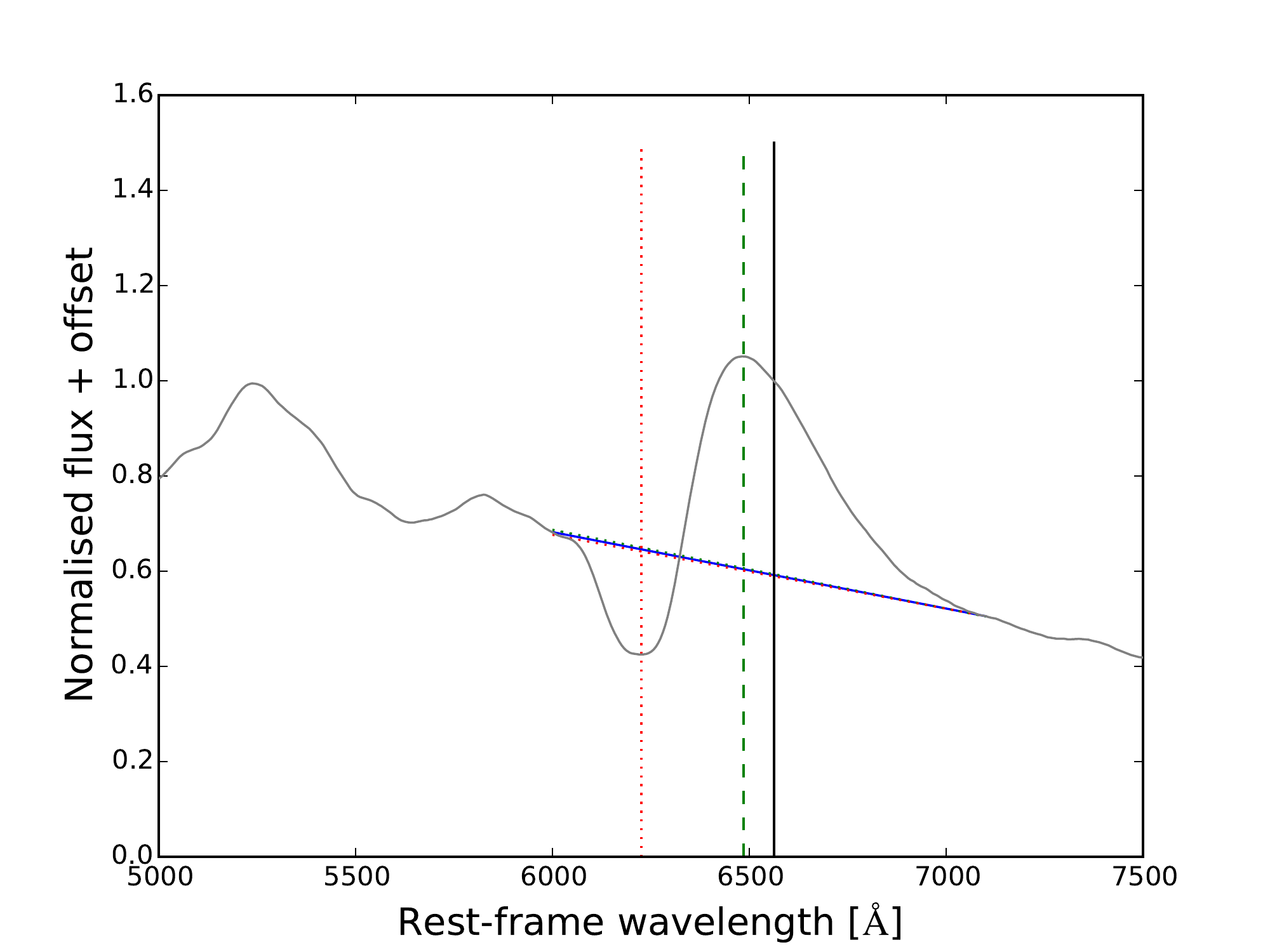}
	\caption{A demonstration of our $f_\mathrm{em}/f_\mathrm{abs}$ process. A smoothed and normalised spectrum is fit with a short linear continuum and the difference between the flux of the continuum and flux of the base (absorption) and peak (emission) is calculated. the ratio $f_\mathrm{em}/f_\mathrm{abs}$ is returned as an absolute value. Errors are calculated by allowing the limits of the continuum to vary over 40 \AA . }
	\label{fig:demoratio}
\end{figure}

We define a value $f_\mathrm{em}/f_\mathrm{abs}$, which is the ratio of the maximum intensity of the emission peak, relative to the local continuum, to the minimum intensity of the absorption trough. The local pseudo-continuum is calculated by using a linear function fitted to two points either side of the features and the errors on the ratio calculated by allowing these points to vary by 40 \AA\ as shown in Figure~\ref{fig:demoratio}. $f_\mathrm{em}$ and $f_\mathrm{abs}$ are then calculated by taking the absolute value of the continuum-subtracted maximum/minimum intensity. This value is derived from the P-Cygni profile and so is a useful measure of the extent of the envelope and the distribution of material within it. Very H-rich SNe have dominant emission peaks due to an extended H envelope.

We then calculate the mean of this value for early-time up to $t_\mathrm{max}$ but because spectral coverage is not so ordered as to provide one spectrum per day for every SN we interpolate the value of $f_\mathrm{em}/f_\mathrm{abs}$ between an early epoch, usually defined as the earliest possible spectrum, to the spectra around $t_\mathrm{max}$. We then take a value for each day and divide by the total number of days to give a pre-peak $\left<f_\mathrm{em}/f_\mathrm{abs}\right>$ and repeat this method for $EW$. On this basis we aim to smooth out any weighting due to clustering of spectral observations and return a robust $\left<f_\mathrm{em}/f_\mathrm{abs}\right>$ and $\left<EW\right>$ that is representative of the early data. A final point is that we omitted the early phases of SNe that displayed indications of shock-breakout because the lack of lines in these SNe is reflective of high ionisation and not low neutral \ion{H}{} opacity.

Typically the errors are relatively small, with the largest errors appearing when the emission component is weak. There is a note of caution with regard to this measurement in that the Doppler shifted absorption of \ion{He}{I} $\lambda$ 6678 can be present as a ``v'' shaped feature on top of the H$\alpha$ emission. The effect can be to reprocess photons in the region between 6563 \AA\ and the absorption feature effectively flattening the region around the emission peak.

\subsection{Other Balmer lines}
\begin{figure}
	\centering
	\includegraphics[scale=0.4]{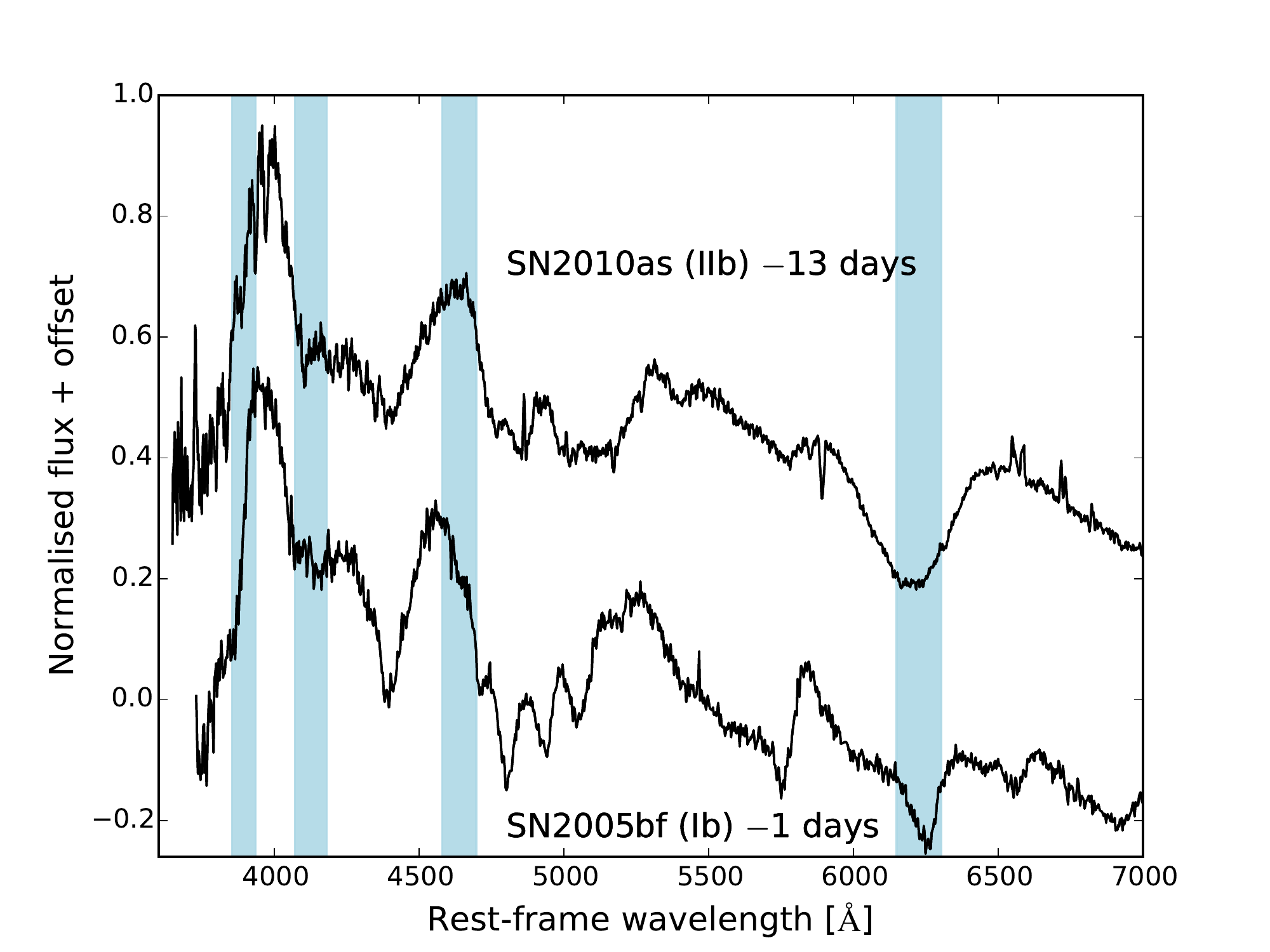}
	\caption{Example spectra of Type IIb SN 2010as  and Type Ib SN 2005bf showing the regions (blue) occupied by H$\alpha-\delta$ at $v\sim 14,000 - 17,000$ km s$^{-1}$. In SN 2010as, at $13$ d before maximum, there is no clear indication of the H lines other than H$\alpha$ and nor do these lines appear at later times. In SN 2005bf ($1$ d before the first peak) there is a notch on the emission profile which may be an indication of H$\beta$, but there is no evidence of H lines at wavelengths shorter than this and this feature disappears after a few days.}
	\label{fig:hbeta}
\end{figure}

Lines of the Balmer series other than H$\alpha$ are normally a prominent feature of Type IIb SNe, however the identification of these lines in some of the SNe IIb is not clear. In Figure~\ref{fig:hbeta} it is shown that Type IIb SN 2010as lacks prominent H$\beta$ and H$\gamma$ lines whereas Type Ib SN 2005bf shows hints of H$\beta$ in its spectra. Similar features are not seen in other SNe in the sample because we are examining a regime where H$\beta$ must be strong enough to have some influence on the spectrum but not so strong that it forms a line which, along with S/N, limits numbers. Without the higher Balmer lines, the certainty of H being present in the ejecta diminishes significantly. This is why it is useful to trace the evolution of these lines and the properties of the H$\alpha$ profile across the range of He-rich SNe as it demonstrates the pathway from H-rich to H-poor ejecta. 

Further discussion on the nature of the $6200$ \AA\ feature which we have broadly attributed to H$\alpha$, and the presence of H in the nebular phase, can be found in Appendix~\ref{sec:app}.

\section{He-rich SNe - Reclassification} \label{sec:4}
\begin{figure}
	\centering
	\includegraphics[scale=0.4]{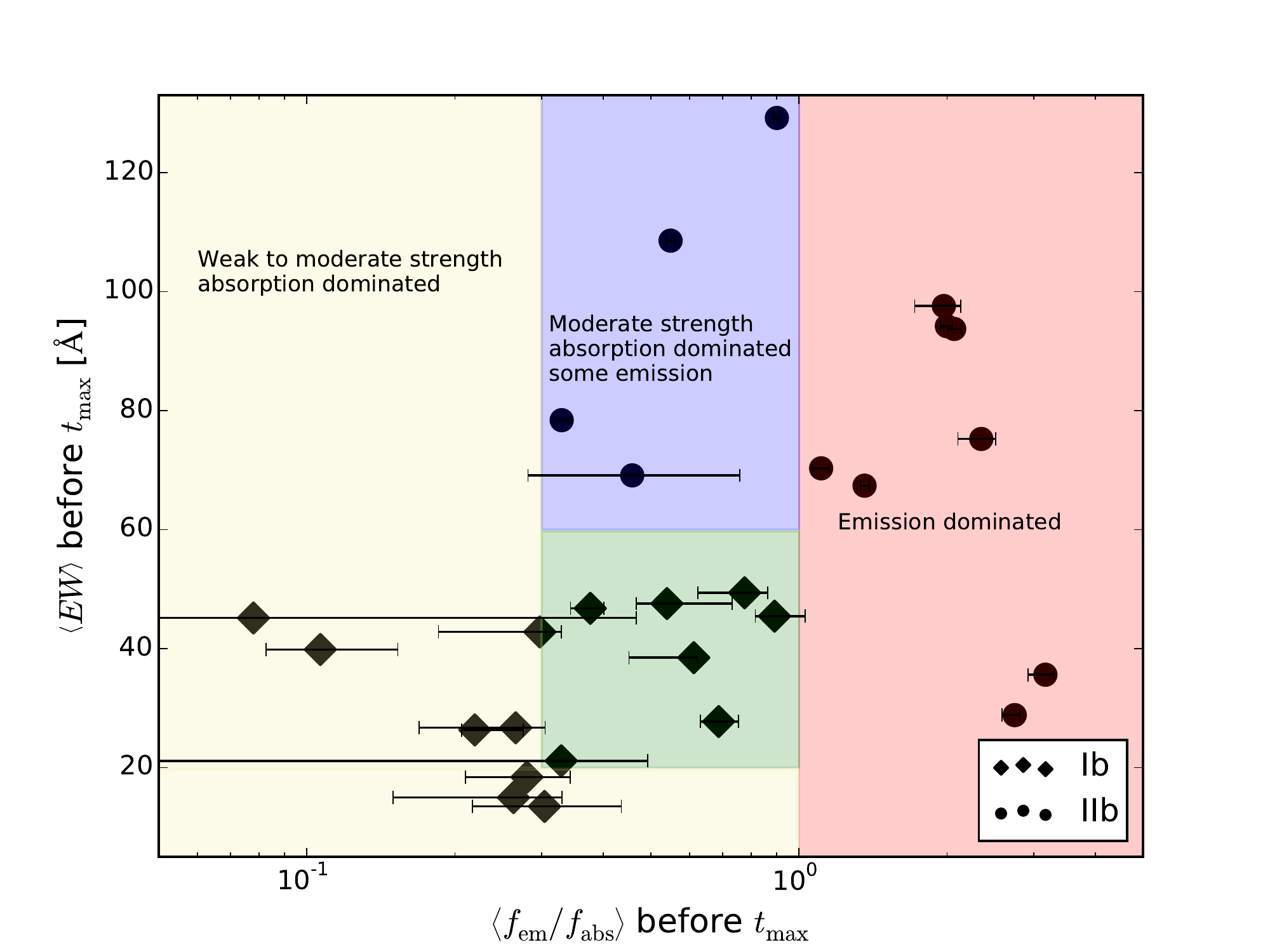}
	\caption{$\left<EW\right>$ of H$\alpha$ as a function of $\left<f_\mathrm{em}/f_\mathrm{abs}\right>$ before peak, with this ratio being presented in logarithmic space for clarity. The plot is segmented to show that some SNe (red) have a flux excess $\left<f_\mathrm{em}/f_\mathrm{abs}\right> >1$, so the emission component dominates the absorption component, these SNe also tend to have moderate to large $EW$ values. The group in the blue region have $0.3< \left<f_\mathrm{em}/f_\mathrm{abs}\right> <1.0$, denoting absorption dominance but they also have strong lines with $\left<EW\right> > 60$ \AA . Below this is a group (green) with similar line profiles but overall weaker line strength with $20 < \left<EW\right>  < 60$ \AA . Supernovae with either very absorption dominated line profiles and/or weak lines are found in the yellow region. The SNe appear to split between SNe Ib occupying the green and yellow regions while the SNe IIb are in the red and blue regions, because they are classified by similarity to reference spectra. This figure indicates there is a transition from Type IIb to Type Ib SNe through the blue and green regions as these SNe are separated by line strength. The distinct separation between SNe Ib and IIb is due to the ``race to the middle ground'' that occurs due to classification by spectral similarity. }
	\label{fig:EWratio}
\end{figure}

This work has shown the H$\alpha$ region provides a continuum of line profiles and strengths for those SNe canonically classified as Type IIb (H-rich) and Type Ib (H-poor). We seek to re-evaluate these SNe in line with the degree of envelope stripping using the H$\alpha$ line as the discriminator. 
In Figure~\ref{fig:EWratio} we plot $\left<EW\right>$ as a function of $\left<f_\mathrm{em}/f_\mathrm{abs}\right>$ for spectra up to the time of maximum light, giving a comparison of line strength against the line profile. Examination of the figure shows that the IIb SNe take the most extreme values of $\left<f_\mathrm{em}/f_\mathrm{abs}\right>$ and $\left<EW\right>$ but there are regions of some similarity. While it may be compelling to view the distribution as evidence of two clear classes the separation is actually an artefact of the classification process. In this case there is a ``race to the middle ground'' as over time SNe are classified due to spectral similarity with earlier SNe and their spectra become reference spectra \citep[e.g.,][]{Modjaz2015} which broadens the definition until the two labels meet at some position in the middle. This appears to be linked to the visibility of the Balmer lines at $\lambda$ shorter than \Ha. The consequence of this is that the distinction between H-rich and H-poor SNe does not reflect the continuum of H abundances.

Using these characteristics as a way of defining the H$\alpha$ feature we propose the following classification system for the SNe listed in Table~\ref{tab:groups}. 

\begin{figure*}
	\centering
	\includegraphics[scale=0.9]{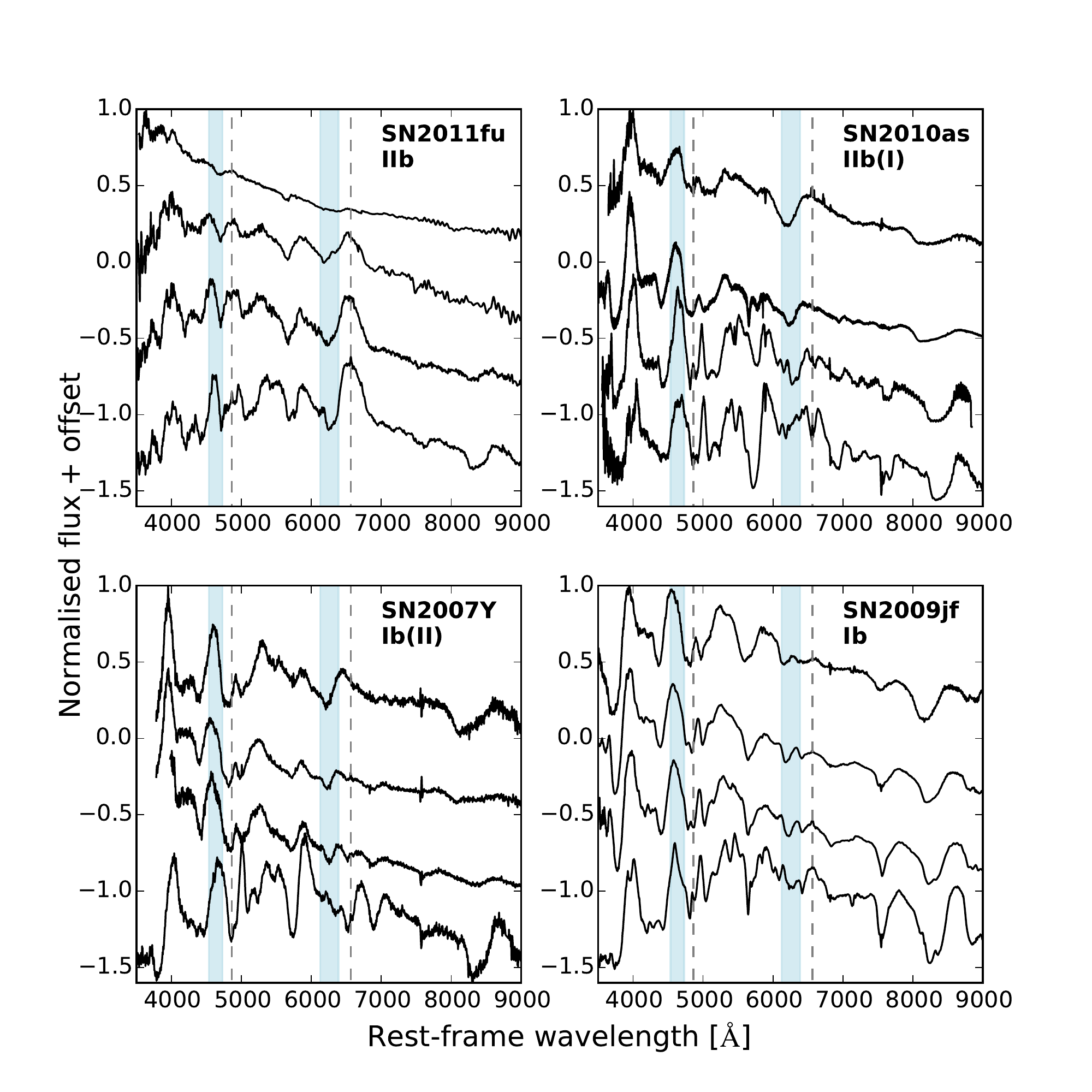}
	\caption{The evolution of representative spectra for SNe in the groupings related to the H$\alpha$ $\left<EW\right>$/ $\left<f_\mathrm{em}/f_\mathrm{abs}\right>$ plane over the course of a month from the earliest spectrum available with an approximate cadence of 7 days. The dashed lines are placed at the rest wavelength of H$\alpha$ and H$\beta$ while the blue regions signify $8,000 < v_\mathrm{H} < 20,000$ km s$^{-1}$ and define the region of interest. (Top left) H-rich IIb SN 2011fu shows many defining characteristics of SNe IIb; a hot shock-breakout phase early on followed by very prominent H lines with a strong H$\alpha$ emission component. (Top right) SN 2010as is an example of the IIb(I) group, clearly the H$\alpha$ profile is weaker than in SN 2011fu and decays strongly over time. The later spectra are \ion{He}{I} and \ion{Fe}{II} dominated blueward of H$\alpha$ with no sign of H$\beta$. (Lower left) SN 2007Y, a member of the Ib(II) group, shows indications of a moderate to weak H$\alpha$ line but there is ambiguity due to the presence of \ion{He}{I} 6678, which cuts in to the H$\alpha$ emission region. The early line profile shows some similarities to that of SN 2011fu and SN 2010as. (Lower right) The Type Ib SN 2009jf displays a weak feature around $6200$ \AA\ which is very quickly overwhelmed by \ion{He}{I} 6678. The presence of any amount of \ion{H}{} is unclear here and it may be possible to entirely attribute the line to \ion{Si}{II}.}
	\label{fig:examples}
\end{figure*}

\subsubsection{SNe IIb}
These SNe have a pre-peak $\left<f_\mathrm{em}/f_\mathrm{abs}\right> > 1$ indicating an emission dominated P-Cygni profile. These SNe are H-rich and all display a prominent series of Balmer lines in their spectra well past peak. In our sample every SN that falls into the IIb category was originally classified as a Type IIb SN and remains so here. In Figure~\ref{fig:examples} we show SN2011fu as an example for this group and demonstrate its evolution over the course of a month.

\subsubsection{IIb(I)}
The IIb(I) group are defined by $0.3 < \left<f_\mathrm{em}/f_\mathrm{abs}\right> < 1$ and $\left<EW\right> >60$ \AA . They typically show strong H line profiles pre-peak but these lines weaken greatly over time. Their mean H$\alpha$ P-Cygni profile is dominated by the absorption component and in some cases the remaining Balmer series lines can be weak or even unidentifiable. Of our sample all these SNe were originally classified as Type IIb and we use SN 2010as as a representative for this group in Figure~\ref{fig:examples}.

\subsubsection{Ib(II)}
The Ib(II) group shows approximately half the H$\alpha$ line strength of the IIb(I) group, $20< \left<EW\right> <60$ \AA , but occupy the same $ \left<f_\mathrm{em}/f_\mathrm{abs}\right>$ range. There is no obvious indication of the Balmer series of lines above H$\alpha$ here. The SNe that are found in this region were all originally classified as Type Ib, however, analysis suggests that the SNe in this group have some amount of H in their ejecta but that it is relatively small compared to the IIb and IIb(I) groups, and so they represent the transition point between H-rich and H deficient SNe. An example of this group, SN 2007Y, is shown in Figure~\ref{fig:examples}.

\subsubsection{Ib}
The final group is defined by typical SNe Ib. The SNe have weak absorption features around $6200$ \AA\ but the sheer lack of strength in emission is such that $ \left<f_\mathrm{em}/f_\mathrm{abs}\right> <0.3$. The weak feature, in conjunction with \ion{He}{I} line scattering in the same region, results in an ambiguous identification of this line. The $~6200$ \AA\ line could be due to a very small mass of \ion{H}{} or due to \ion{Si}{II}. We use SN 2009jf as the representative of this group and plot it in Figure~\ref{fig:examples}.

\begin{table}
	\centering
	\caption{The grouping of He-rich SNe with respect to H$\alpha$ profile }
	\begin{tabular}{cccc}
	IIb & IIb(I) & Ib(II) & Ib \\
	SN1993J & SN1996cb & SN2007Y & SN2009jf \\
	SN2011fu & SN2008ax & SN1999ex & iPTF13bvn \\
	SN2003bg & SN2006el & SN2008D & SN2004gq \\
	SN2011ei & SN2010as & SN2007kj & SN2009er \\
	SN2011dh&  & SN1999dn & SN2005hg \\
	SN2006T &  & SN2005bf & SN2009iz \\
	SN2011hs & & SN2006ep & SN2006lc \\
	SN2008bo & & SN2007uy &\\
	\hline
	\end{tabular}
	\label{tab:groups}
\end{table}

\section{He-poor SNe: Type Ic} \label{sec:5}
\begin{figure}
	\centering
	\includegraphics[scale=0.4]{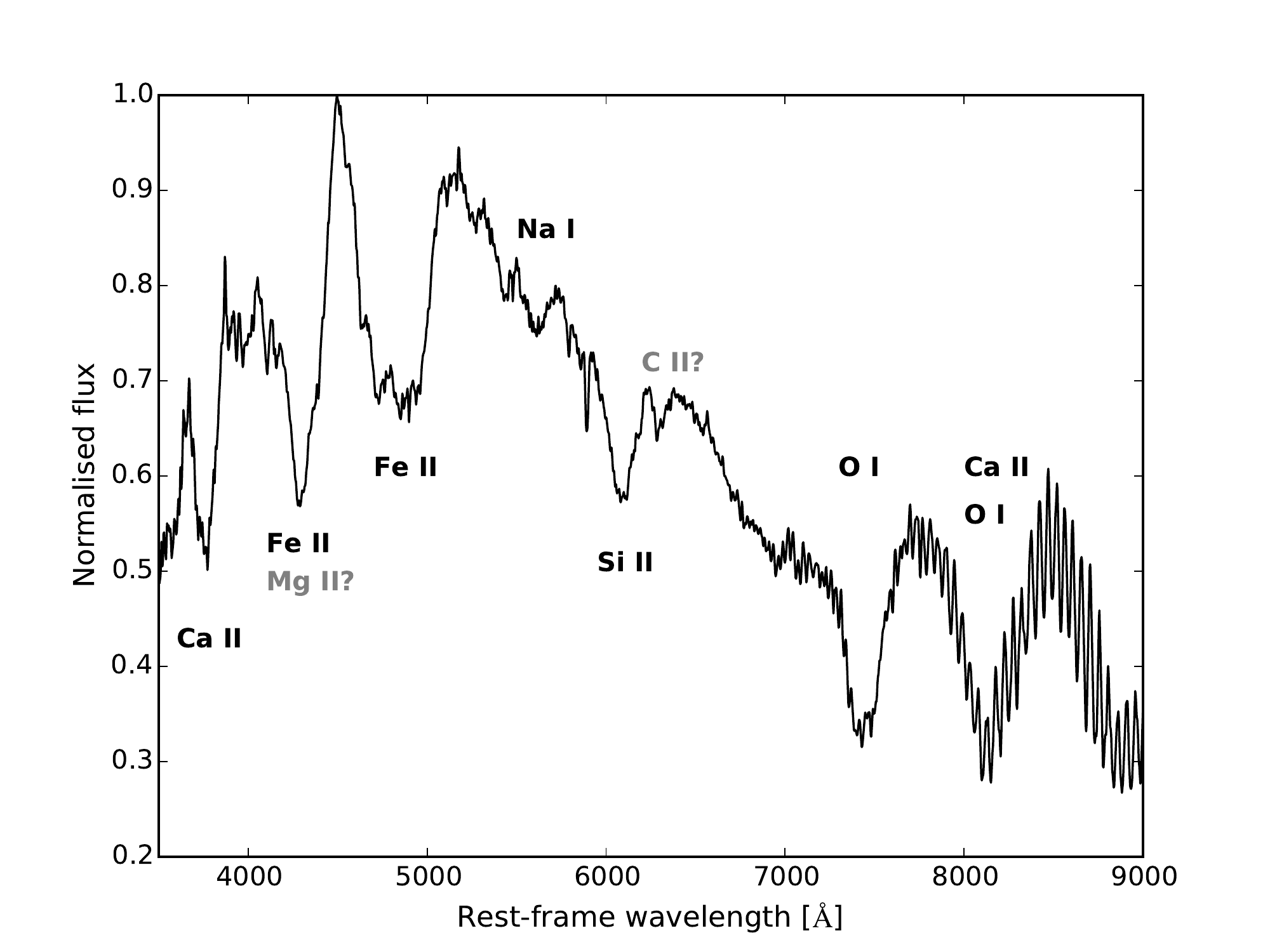}
	\caption{A spectrum of Type Ic SN 2004aw \citep{Taubenberger2006} showing the ions that are responsible for the most prominent absorption features in a typical Type Ic SN. As with any expanding atmosphere, the lines are blue shifted in relation to the velocity of the material that is responsible for the line forming regions, in this case by $10,000-15,000$ km s$^{-1}$. The diversity of type Ic spectra is principally a product of the velocity and degree of blending of these lines. }
	\label{fig:Ic}
\end{figure}

The initial analysis described in Section~\ref{sec:2} showed that there was a clear division between the He-rich SNe and the He-poor SNe; those of Type Ic \citep[for an innovative investigation into the presence of He in SNe Ic see][]{Modjaz2015}. Unlike the He-rich SNe, in which the spectra are dominated by H and or He lines, SNe Ic are dominated by \ion{Fe}{II}, \ion{Si}{II}, \ion{O}{I}, \ion{Ca}{II}, and possibly \ion{Mg}{II} and \ion{C}{II} as shown in Figure~\ref{fig:Ic}.

\begin{figure}
	\centering
	\includegraphics[scale=0.4]{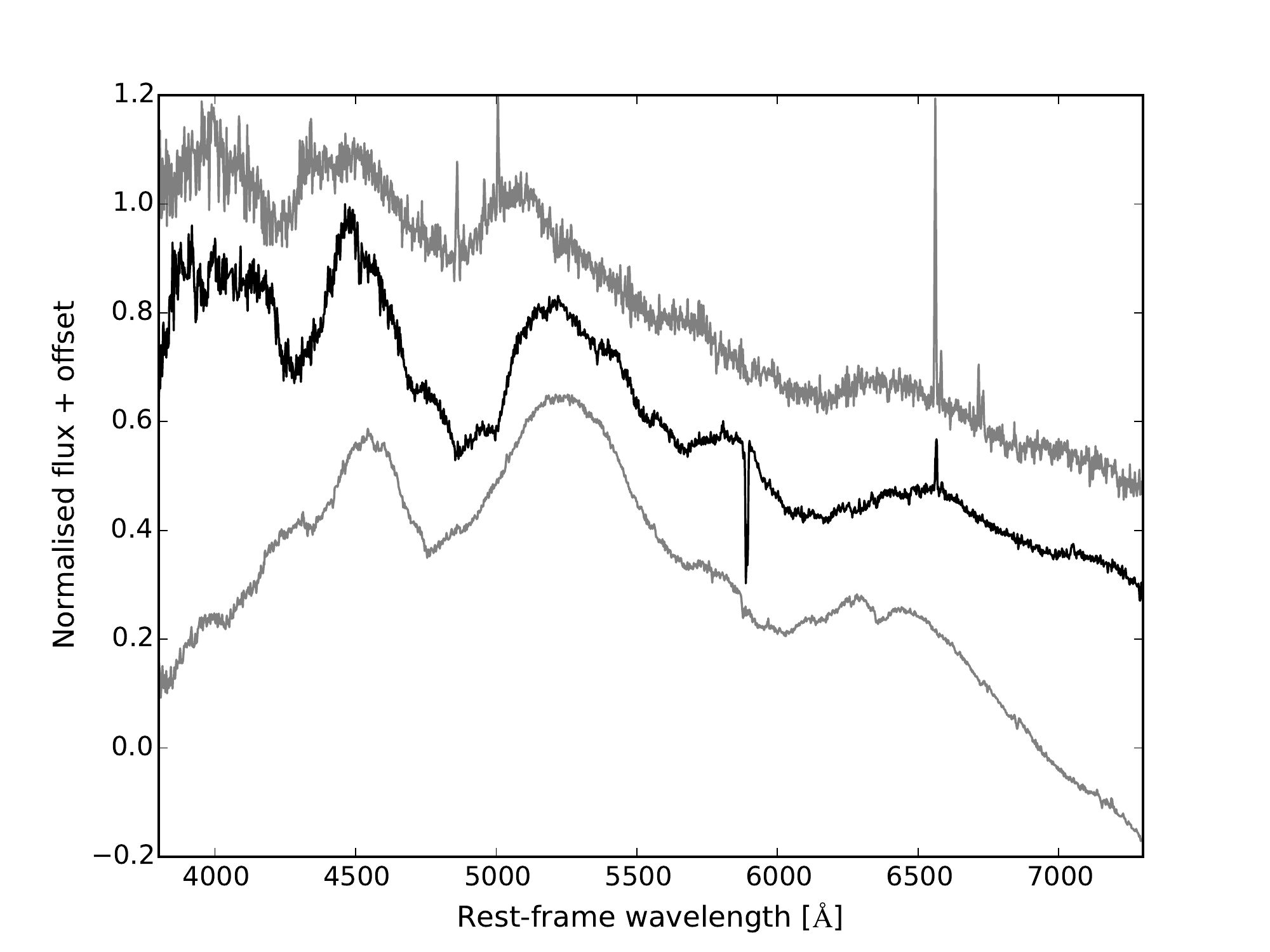}
	\caption{A comparison of the first spectrum of SN 1994I a few days after explosion (black) with Ic-BL SNe SN 2003jd (top) and SN 2002ap (bottom). The spectrum of SN 1994I shows similarities with both SNe and although it is considered a typical SN Ic and the lines narrow shortly after this time, it does demonstrate the subjective nature of the Ic-BL categorisation.}
	\label{fig:bl}
\end{figure}

Here SNe Ic are analysed within the context of line broadening because this is the defining characteristic used to classify them. If the SN shows ``narrow'' lines \cite[e.g. SN 1994I][]{Filippenko1995} it will be labelled as a Type Ic, while if the SN has ``broad'' lines it will be labelled Ic-BL. Unfortunately, the definition of ``broad-lined'' is subjective and tells us little about the properties of the spectra other than the SN has broader lines than SN 1994I. 
To confuse matters, the earliest spectrum of SN 1994I also showed shallow, broad lines before giving way to narrower features a day later, as shown in Figure~\ref{fig:bl}. Had SN 1994I exploded a decade later then it is probable that it would have been classified as Ic-BL, at least initially. With more data it is now possible to quantify the degree of line broadening and provide useful information in the classification scheme.

\subsection{Methodology - Counting features}
We adopt a straight forward method to analyse the spectra of SNe Ic by counting the number of absorption features $N$ that are visible. By doing this we take a measure of the degree of line blending which can then be related to $M_\mathrm{ej}$, $E_\mathrm{k}$, and the properties of their light curves. The intention is that this procedure should be easily replicable and provide information about the explosion characteristics but without the need for more complex analytic methods or modelling.
 
The first step is to note the constraints this method must work under, thus we must account for the range of wavelengths covered by different spectrographs and their respective S/N and ``a feature'' must be defined in such a way that is not ambiguous. With this in mind we take a core wavelength range of $4000 - 8000$ \AA\, as this is within the range of most spectrographs. We then define a set of characteristic features within this range and one outside, as defined by the lines given in Table~\ref{tab:lines}. We include the \ion{Ca}{II} NIR triplet because it is often, but not always, covered by spectroscopy and its presence is ubiquitous in SE-SNe. Furthermore we can infer properties of this line by observing the \ion{O}{I} 7774 triplet as well as the \ion{Ca}{II} H$\&$K lines if the spectrum extends sufficiently far to the blue. For the former case the shape of the line will indicate a lack of blending if the line is sufficiently narrow and the spectrum extends far enough to capture the rise of the red-wing of the line. For the latter the velocity of \ion{Ca}{II} can be measured, given that velocities of less than $\sim 27,000$ km s$^{-1}$ are unlikely to result in line blending. The \ion{C}{II} 6580 \AA\ line, which is normally identifiable, is not counted because it is relatively weak and is close in velocity space to the much stronger \ion{Si}{II} line.

Next, the lines in the spectra are examined noting which are blended and which are absent, and calculating the line velocity from the bottom of the absorption profile. Velocities are measured up to 30,000 km s$^{-1}$, although values in excess of this are difficult to measure as such velocities also lead to line blending. 

Note that by attributing a feature to one of these lines we do not suggest it is a positive identification of that element. This is especially relevant to the region occupied by \ion{Na}{I} D and \ion{Si}{II} 6347 as several unidentified features can appear here. Thus we count a maximum of two features in this region. We define a feature as being a significant change in the spectrum that cannot be attributed to noise. Finally, we do not count all the visible features in a spectrum because each feature should be identifiable in other SNe spectra and the evolution of these features should be easily traceable.

\begin{table}
	\centering
	\caption{Prominent spectral lines used to determine $N$}
	\begin{tabular}{ll}
	Ion & $\lambda$/ [\AA ] \\
	\ion{Fe}{II} & 4924\\
	\ion{Fe}{II} & 5018\\
	\ion{Fe}{II} & 5198\\
	\ion{Na}{I} & 5895\\	
	\ion{Si}{II} & 6347\\
	\ion{O}{I} & $\sim$7774 triplet\\
	\ion{Ca}{II} & NIR triplet\\
	\hline
	\end{tabular}
	\label{tab:lines}
\end{table}

\subsubsection{Noise and other sources of contamination}
\begin{figure}
	\centering
	\includegraphics[scale=0.4]{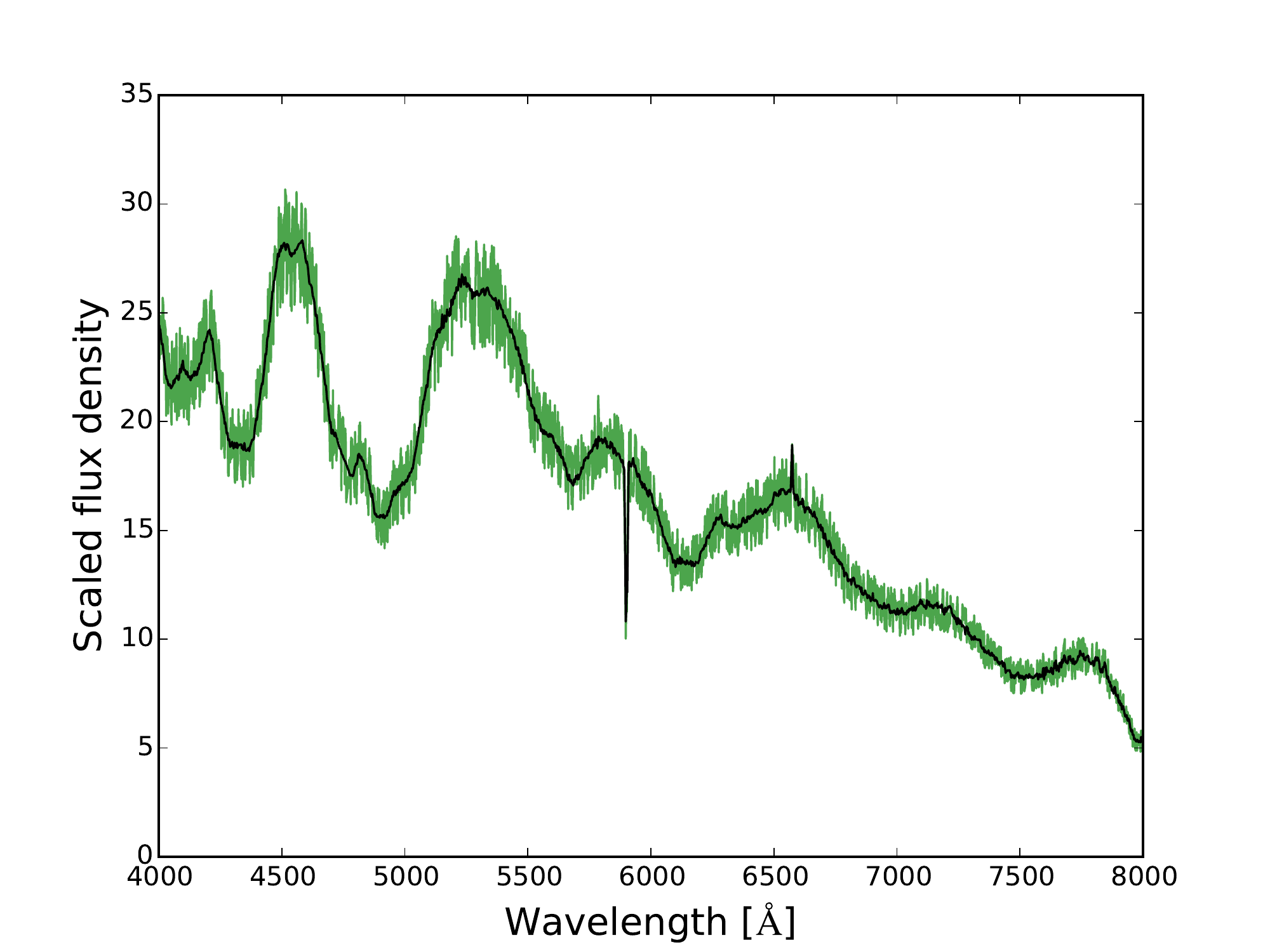}
	\caption{An example the effect of noise on spectra. Here a spectrum of SN 1994I (black line) has had noise artificially inserted by allowing the actual flux values to vary by $\pm{10}\%$ (green). While most of the features appear unaffected the region $\sim4700\AA$ corresponding with the \ion{Fe}{II} 4924 and 5018 lines becomes ambiguous while in the original spectrum it is clear they are blended with each other but not with the \ion{Fe}{II} 5169, 5198, 5235 series. The clean spectrum would return $N=6$ while the noisy spectrum would return $N=5$, because of this we would define the noisier spectrum as $N>5$, indicating the lower limit. If cleaner spectra are available around the same time then this limit can replaced by an actual value}
	\label{fig:noise}
\end{figure}

The method of counting features is sensitive to the resolution and noise of the individual spectra. Noisy spectra make feature identification difficult particularly if there are groups of lines in close proximity to each other (e.g., the \ion{Fe}{II} lines around 5000 \AA ) or in the case where features become very broad and shallow, such as with hypernovae (e.g., SN 2002ap and SN 1998bw). We provide an example of how noise affects the spectra in Figure~\ref{fig:noise}. Additionally, contamination from emission by other processes (e.g., host-galaxy, GRB afterglow) must also be considered because such processes can mask the SN spectrum leading to an absence of certain absorption lines which results in a flat region of the spectrum. This can be a serious problem for GRB/XRF SNe because the afterglow flux follows a power law decay in temporal and wavelength space which means that the contribution to the total flux from this component is stronger during the critical early phases.

To deal with this the first option is to check spectra taken shortly before and after the spectrum in question. This may allow a more robust estimate of the lines and their evolution at the epoch in question rather than relying on the noisy spectrum. The second option is to give a lower limit to the number of features. Given that noise primarily affects the \ion{Fe}{II} lines around $5000$ \AA\ this provides an uncertainty of 2 lines. If the noise was excessive enough to mask other lines then the spectrum would be rejected.

\subsubsection{Broad lines and $E_\mathrm{k}$/$M_\mathrm{ej}$}
The importance of $N$ is that it is a measure of the degree line blending and broadening in SNe, which is linked to the intrinsic density profile of the ejecta and which in turn affects $E_\mathrm{k}$. The total kinetic energy $E_\mathrm{k}$ of the ejecta is a measure of the energy of the explosion while ejecta mass, $M_\mathrm{ej}$, can give indications as to the zero-age main sequence (ZAMS) mass of the progenitor whereby a larger $M_\mathrm{ej}$ correlates with a greater $M_\mathrm{ZAMS}$ \citep{Nomoto1994}. $E_\mathrm{k}$/$M_\mathrm{ej}$ is the specific kinetic energy, giving the kinetic energy per unit mass, and as $E_\mathrm{k}$ is dominated by high velocity material it is also related to the explosion mechanism. This is due to the fact that the total $E_\mathrm{k}$ is sensitive to the shape of the outer part of the ejecta density profile because a shallower density profile adds mass, and therefore opacity, at higher velocities. The addition of a small amount of mass a $v > 25,000$ km/s can add $\approx 10^{52}$ erg to the kinetic energy \citep[see, for example,][]{Mazzali2000,Mazzali2013,Nakamura2001}.

\section{Results} \label{sec:6}
\begin{figure*}
	\centering
	\includegraphics[scale=0.7]{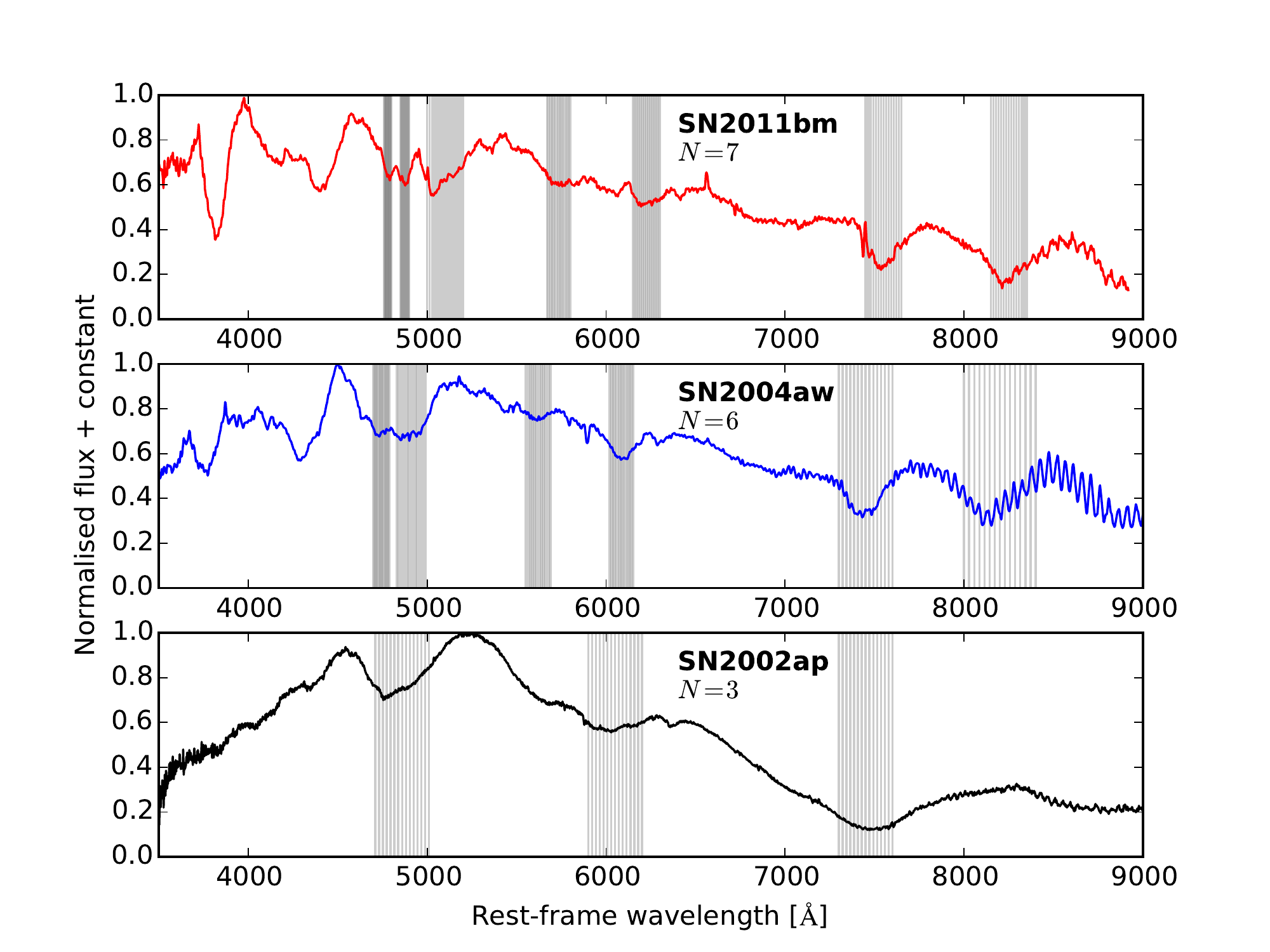}
	\caption{(Top) A spectrum of SN 2011bm with $N=7$, as can be seen the \ion{Fe}{II} 4924 and 5018 lines are clearly unblended. Because of the limited wavelength range of many spectrographs we do not count anything blueward of the \ion{Fe}{I} 4924 line, however, if present the \ion{Ca}{II} H$\&$K lines can be used to limit the velocity of the \ion{Ca}{II} shell. The absorption feature directly blueward of the \ion{Fe}{II} 4924 line is a further series of \ion{Fe}{II} lines. (Middle) SN 2004aw shows blended \ion{Fe}{II} lines although the rest of the designated features are unblended. (Bottom) A spectrum of SN 2002ap indicating severe line blending. The \ion{Fe}{II} lines are completely blended as are O and Ca. The \ion{Si}{II} 6347 feature is broad, however it is not blended with the \ion{Na}{I} D line, which itself is not clearly apparent. It may be appearing around $\sim5700$ \AA\ but is not counted here.}
	\label{fig:featuressample}
\end{figure*}

Figure~\ref{fig:featuressample} shows the results found when applying this method to three SNe covering narrow to broad absorption features. The need for good S/N is clearly seen when considering the \ion{Fe}{II} regions.

$N$ evolves differently over time for each SN but during the period before peak, and for some days afterwards, $N$ increases as a function of time and never decreases. At later times the identification of features becomes more difficult as the continuum flux decreases, especially in the blue near the \ion{Fe}{II} lines, and the density of the outer ejecta decreases sufficiently to significantly reduce opacity in some lines, in addition to this the structure of the spectra becomes more complex as low velocity elements become more prominent. The evolution of $N$ before this epoch is towards 6 and 7 in most cases but in some situations (e.g. SN 1998bw, PTF10vgv \citep{Corsi2012}) significant line blanketing around $5000$ \AA\ masks \ion{Fe}{II} lines in this region, restricting the value of $N$.

\begin{figure*}
	\centering
	\includegraphics[scale=0.7]{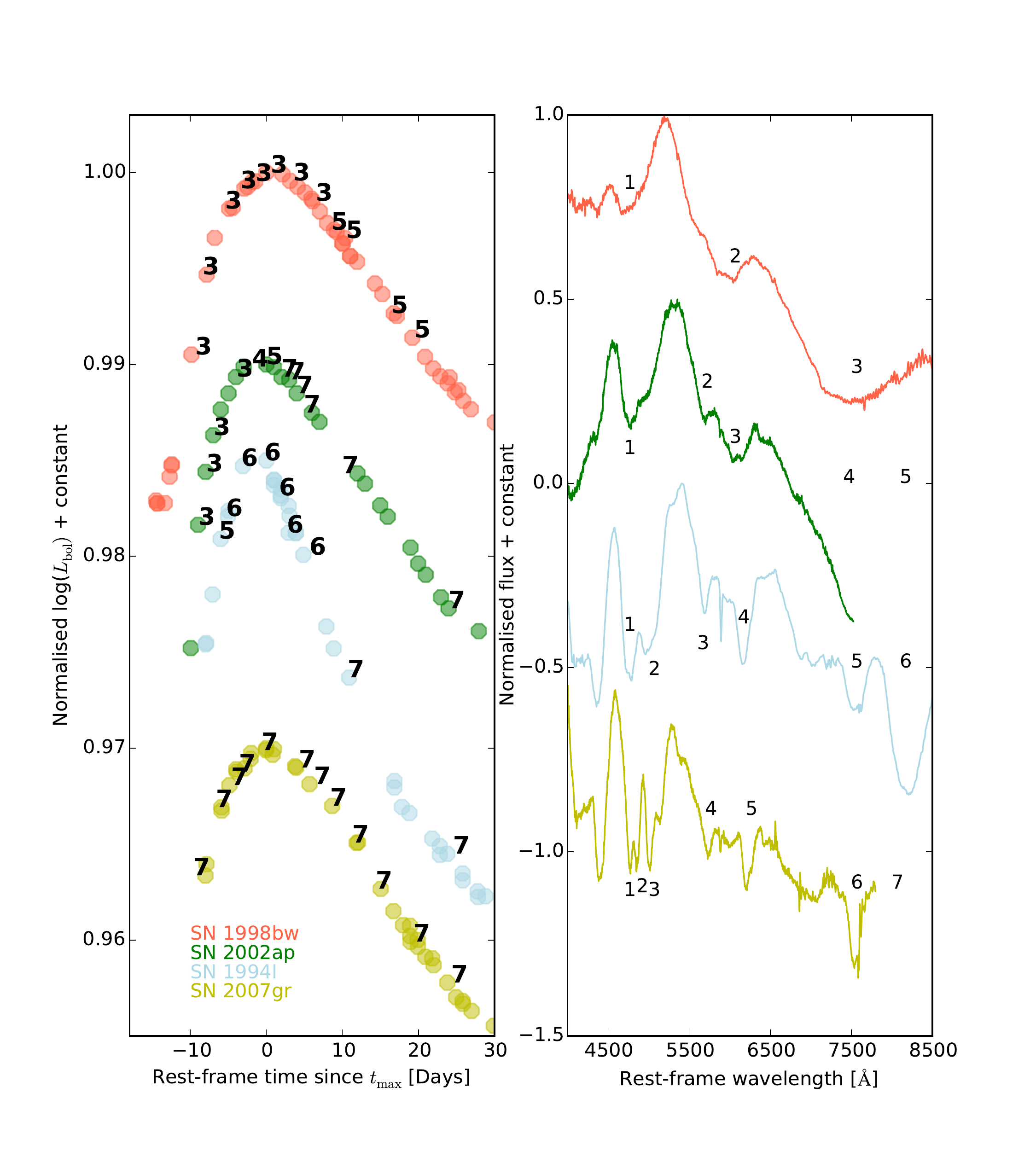}
	\caption{(Left) The normalised and offset bolometric light curves of our reference SNe. The number of features $N$ in the spectra is shown as a function of epoch, the evolution of $N$ is not entirely predictable as shown by SN 2002ap where $N=3$ before maximum but rapidly evolves to $N=7$ shortly after maximum while SN 1994I takes much longer to reach $N=7$. (Right) The maximum light spectra of the reference supernovae, the number of features counted and their location. The \ion{Ca}{II} NIR triplet is present in SN 2007gr at this epoch however the spectrum does not extend sufficiently far to the red to show it, while in SN 2002ap there is enough evidence in spectra around this epoch to suggest that the lines are in the process of separating.}
	\label{fig:bigplot}
\end{figure*}

As an example of the evolution of $N$, Figure~\ref{fig:bigplot} shows the $t_\mathrm{max}$ spectrum and the evolution of $N$ in relation to the light curve for four SNe in our sample; SN 1998bw \citep{Patat2001}, SN 2002ap \citep{Mazzali2002,Foley2003}, SN 1994I \citep{Richmond1996,Filippenko1995,Sauer2006}, and SN 2007gr \citep{Valenti2008,Mazzali2010}. There is clear diversity in the shape and temporal characteristics of each SN and the evolution of $N$ is not predictable or uniform although $N$ never decreases with time. This latter behaviour allows upper limits on $N$ to be placed at times earlier than the first spectrum.

\subsubsection{$N$ at $t_{-1/2}$, $t_\mathrm{max}$, and $t_{+1/2}$} \label{sec:N}
$N$ can be evaluated as a function of time but because of the discrete values of $N$ and the different temporal characteristics of each SN we consider just the value of $N$ at comparative epochs in the evolution of the LC.
The time of bolometric maximum $t_\mathrm{max}$ is used as a fiducial $t_0$. We also consider two other reference times, the first is \trise, which is the time taken for the light curve to rise from $L_\mathrm{p}/2$ to peak luminosity $L_\mathrm{p}$, $t_{-1/2}$. The second is \tdecay, the time taken for the light curve to decay from $L_\mathrm{p}$ to $L_\mathrm{p}/2$. These times are used to investigate the number of features in the spectra scaled to a similar epoch in the evolution of the light curve. Spectra are selected within a bin around $\pm{2}$ days for $t_{-1/2}$, $\pm{4}$ days for $t_\mathrm{max}$ and $\pm{6}$ days for $t_{+1/2}$, where the windows reflect the fact that a day represents a greater proportion of the evolved time earlier in the light curve.

In some cases there was sufficient photometry to measure $t_{-1/2}$ for individual bands but insufficient photometry to construct a bolometric LC. To compensate for this we took the SNe with bolometric $t_{-1/2}$ and investigated how the multi-band $t_{-1/2}$ varied in comparison. It is found that $t_{-1/2}$ for the {\it gVR} bands were generally within two days of the bolometric $t_{-1/2}$, and for most SNe it was closer to one day. Thus, for SNe which lacked a bolometric $t_{-1/2}$ but had earlier photometry we used the $t_{-1/2}$ for {\it VRg}, prioritised by their respective order.  

Table~\ref{tab:risefeat} gives the line velocities and overall $N$ of SNe with spectra at $t_{-1/2}$ while Table~\ref{tab:peakfeat} gives the same for spectra at $t_\mathrm{max}\pm4$ days. Finally, the spectral velocities at $t_{+1/2}$ are given in Table~\ref{tab:peakdecay}; we do not list $N$ here because the degree of blending in the \ion{Fe}{II} region between $4500 - 5000$ \AA\ is ambiguous owing to the decaying strength of these lines in some of the spectra. If the S/N is sufficiently good then in most cases $N$ takes a value of 6 or 7. The post peak evolution of He-poor SNe spectra is discussed in Section~\ref{sec:pp}.

\begin{table*}
	\centering
	\caption{$N$ and estimated velocities in SNe Ic at $t_{-1/2}$}
	\begin{tabular}{lcccccccccc}
	SN & Type & Epoch & \ion{Fe}{II} 4924 & \ion{Fe}{II} 5018 & \ion{Fe}{II} 5198 & \ion{Na}{I} & \ion{Si}{II} & \ion{O}{I} & \ion{Ca}{II}& $N$ \\
	& & [days] & [km s$^{-1}$] & [km s$^{-1}$] & [km s$^{-1}$] & [km s$^{-1}$]& [km s$^{-1}$] & [km s$^{-1}$] & [km s$^{-1}$] & \\
	1994I & Ic & -5.26 & \multicolumn{2}{c}{16,000  } & 16,000 & 11,000 & 11,000 & 11,000 & 17,000 & 6 \\
	1998bw & GRB-SN & -9.33 & \multicolumn{3}{c}{Blended} & \multicolumn{2}{c}{Blended } & \multicolumn{2}{c}{Blended } & 3 \\
	2002ap & Ic-BL & -6.98 & \multicolumn{3}{c}{Blended} & Flat & 23,000 & \multicolumn{2}{c}{Blended } & 3 \\
	2004dn & Ic & -9.52 & \multicolumn{3}{c}{Blended?} & 10,000? & 9,000 & 10,000 & 15,000  & >5$^{a}$ \\
	2004fe & Ic & -7.8 & \multicolumn{2}{c}{13,000 } & 13,000 & 9,000 & 9,000 & 11,000 & 16,000 & 6 \\
	2006aj$^\ddag$ & XRF-SN & - 5.3 &  \multicolumn{3}{c}{Blended} & Flat & 14,000 & \multicolumn{2}{c}{Flat } & 2 \\
	2007gr & Ic & -8.6 & 11,000 & 11,000 & 11,000 & 10,000 & 9,000 & 10,000 & 14,000 & 7 \\
	2009bb & Ic-(BL?) &-4.3 & \multicolumn{3}{c}{Blended} & 30,000 & 30,000 & 17,000 & 30,000 & 5 \\
	2010bh* & GRB-SN & -5.2 & \multicolumn{3}{c}{Blended} & \multicolumn{4}{c}{Flat} & 1\\
	PTF10vgv & Ic & -7.4 & \multicolumn{3}{c}{Blended?} & Flat & 15,000 & Flat & <10,000 & >3 \\
	2011bm*$_\dag$ & Ic & -14.2 &10,000 & 10,000 & 10,000 & 9,000 & 6,000? & 10,000 & 10,000 & 7  \\
	PTF12gzk* & Ic & -8.5 & \multicolumn{2}{c}{27,000 (Blended) } & 27,000 & 22,000 & 21,000 & 23,000 & 23,000 & 6 \\
	\multicolumn{11}{l}{$^{a}$ N is almost certainly 6 or 7}\\
	\multicolumn{11}{l}{$\ddag$ SN 2006aj is severely contaminated by afterglow and host-galaxy lines}\\
	\multicolumn{11}{l}{*$t_{-1/2}$ estimated from {\it V} or {\it R} band photometry}\\
	\multicolumn{11}{l}{$^\dag$ $t_{-1/2}\sim$ 22 days, this spectrum is the first with decent S/N}\\
	
	\end{tabular}
	\label{tab:risefeat}
\end{table*}

\begin{table*}
	\centering
	\caption{Number of features and estimated velocities in SNe Ic at $t_\mathrm{max}$}
	\begin{tabular}{lcccccccccc}
	SN & Type & Epoch & \ion{Fe}{II} 4924 & \ion{Fe}{II} 5018 & \ion{Fe}{II} 5198 & \ion{Na}{I} & \ion{Si}{II} & \ion{O}{I} & \ion{Ca}{II} & $N$ \\
	& & [days] & [km s$^{-1}$] & [km s$^{-1}$] & [km s$^{-1}$] & [km s$^{-1}$]& [km s$^{-1}$] & [km s$^{-1}$] & [km s$^{-1}$] & \\
	1994I & Ic & 0.7 & \multicolumn{2}{c}{13,000 } & 13,000 & 11,000 & 9,000 & 11,000 & >13,000 & 6 \\
	1997ef & Ic-BL & -3.0 & \multicolumn{3}{c}{20,000}& flat &13,000 &17,000 & 20,000& 4 \\
	1998bw & GRB-SN & 0.58 & \multicolumn{3}{c}{Blended} & X & 15,000 & \multicolumn{2}{c}{Blended } & 3 \\
	2002ap & Ic-BL & 0.0 & \multicolumn{3}{c}{>15,000} & 11,000 & 9,000 & 10,000 & 15,000 & 5 \\
	2003jd & Ic-BL & 0.25 & \multicolumn{3}{c}{>15,000} & 13,000 & 13,000 & >15,000 & - & >4 \\
	2004aw & Ic & 0.3 &\multicolumn{2}{c}{15,000} & 15,000 & 14,000 & 11,000 & 13,000 & 15,000 & 6\\
	2004fe & Ic & 0.0 & \multicolumn{2}{c}{11,000 } & 11,000 & 10,000 & 8,000 & <12,000$^\dag$ & <17,000$^\dag$ & 6 \\
	2005az & Ic & 1.0 & 8,000 & 8,000 & 8,000 & 6,000 & 11,000 & - & - & 7$^\ddag$ \\
	2006aj & XRF-SN & 0.4 &  21,000 & 21,000 & 21,000 & X & 21,000 & 15,000 & 21,000 & 6 \\
	2007gr & Ic & -0.6 & 10,000 & 10,000 & 10,000 & 9,000 & 7,000 & 9,000 & <13,000$^\dag$ & 7 \\
	2009bb & Ic-BL &1.6 & \multicolumn{3}{c}{Blended?} & 22,000 & 20,000 & 19,000 & 24,000 & >5$^{a}$ \\
	2010ah & Ic-BL &-1.5 & \multicolumn{3}{c}{Blended} & X & 18,000 & \multicolumn{2}{c}{Blended} & 3 \\
	2010bh & GRB-SN & 0.0 & \multicolumn{3}{c}{Blended} & 25,000 & 32,000 & - & 37,000 & 4\\
	2011bm & Ic & 0.0 &  - & - & -& - & - & - & - & 7$^\ddag$\\
	PTF10vgv & Ic-(BL?) & 3.4 & \multicolumn{3}{c}{Blended} & Flat & 7,000 & 10,000 & 15,000 & >4 \\
	2012ej & Ic & 2 &  7,000 & 7,000 & 7,000 & 7,000 & 5,000 & - & - & 7\\
	PTF12gzk & Ic & -0.6 & \multicolumn{2}{c}{22,000} & 22,000 & 19,000 & 17,000 & 16,000 & 21,000 & 6 \\
	2016P & Ic-BL &-3.0 &\multicolumn{3}{c}{Host contamination} & 14,000 & Host contamination & - & 16,000 & 6 \\
	2016coi & Ic-BL & 0.0 & \multicolumn{3}{c}{Blended} & 14,000 & 14,000 & \multicolumn{2}{c}{Blended} & 4 \\
	2016iae & Ic & -0.1 & 13,000 & 13, 000 & 13,000 & 11,000 & 9,000 & 11,000 & 16,000 & 7 \\
	\multicolumn{11}{l}{$\dag$ Upper limit derived from earlier epochs}\\
	\multicolumn{11}{p{\textwidth}}{$\ddag$ For SN 2005az we lack spectra redward of $7000$ \AA\ but the number of features is almost certainly 7 as the FeII lines are unblended. For SN 2011bm we are able to give the number of features because they had split before the time of bolometric maximum}\\
	\multicolumn{11}{l}{- denotes that the feature is not covered by spectrum}\\
	\multicolumn{11}{l}{$^{a}$ Contamination by host galaxy lines and depressed flux in blue, likely $N=6$ or 7}\\
	\multicolumn{11}{l}{X denotes that the feature is not visible}\\
	\end{tabular}
	\label{tab:peakfeat}
\end{table*}

\begin{table*}
	\centering
	\caption{The estimated velocities in SNe Ic at $t_\mathrm{+1/2}$. All \ion{Fe}{II} lines are given one velocity, the degree of blending in many cases is unknown at this epoch.}
	\begin{tabular}{lcccccccccc}
	SN & Type & Epoch & \ion{Fe}{II} 4924 & \ion{Fe}{II} 5018 & \ion{Fe}{II} 5198 & \ion{Na}{I} & \ion{Si}{II} & \ion{O}{I} & \ion{Ca}{II}  \\
	& & [days] & [km s$^{-1}$] & [km s$^{-1}$] & [km s$^{-1}$] & [km s$^{-1}$]& [km s$^{-1}$] & [km s$^{-1}$] & [km s$^{-1}$]  \\
	1994I & Ic & 8.7 & \multicolumn{3}{c}{10,000} & 10,000* & 4,000 & 8,000 & 10,000 \\
	1997ef & Ic-BL & 24 & \multicolumn{3}{c}{10,000} &7,000$^\dag$ &4,000 & - & - \\
	1998bw & GRB-SN & 16.4 & \multicolumn{3}{c}{15,000} & 11-13,000 & 6,000 & 5-7000 & 15,000 \\
	2002ap & Ic-BL & 15 & \multicolumn{3}{c}{10-15,000} & 7-11,000 & 4,000 & 6-8,000 & 10-15,000  \\
	2003jd & Ic-BL & 17.9 & \multicolumn{3}{c}{15,000} & 12,000 & 7,000 & - & -  \\
	2004aw & Ic & 20 &\multicolumn{3}{c}{12-15,000} & 7-9,000 & 7,000 & 10,000 & 10,000 \\
	2005az & Ic & 27 &  \multicolumn{3}{c}{9,000} & 7,000 & 13,000 & - & -  \\
	2006aj & XRF-SN & 9 &  \multicolumn{3}{c}{20-23,000} & flat  & 15,000 & 15,000 & 20-23,000 \\
	2007gr & Ic & 14.4 &  \multicolumn{3}{c}{7,000} & 7,000 & 4,000$^\ddag$ & 7,000 &7-8,000\\
	2009bb & Ic-BL &15.5 & \multicolumn{3}{c}{13-25,000} & 16,000 & 12,000 & 13,000 & 16-22,000 \\
	2010bh & GRB-SN & 10 & \multicolumn{3}{c}{Blended} & flat & 30,000 & flat & 30-40,000\\
	2011bm & Ic & 43 &  \multicolumn{3}{c}{6,000}& 6,000 & flat & 7,000 & 6,000\\
	PTF10vgv & Ic-(BL?) & 7.3 & \multicolumn{3}{c}{12,000} & Flat & 5,000 & 10,000 & 12,000 \\
	PTF12gzk & Ic & 26 & \multicolumn{3}{c}{17,000} & 12,000 & 8,000 & 10,000 & 17,000 \\
	\multicolumn{10}{l}{*possibly dominated by \ion{He}{I} 5876}\\
	\multicolumn{10}{p{\textwidth}}{$\dag$ Multi-component base, velocity of red component used}\\
	\multicolumn{10}{p{\textwidth}}{$\ddag$ Ambiguous identification}\\
	\end{tabular}
	\label{tab:peakdecay}
\end{table*}

\subsection{The effect of line-blending on $N$}
The value of $N$ is a discrete number and there is a prevalence for certain values which correspond to particular sets of line blending, which are listed in Table~\ref{tab:blends}. Figure~\ref{fig:featuressample} shows three example SNe with the features highlighted. The effect of line blending and broadening is apparent as one looks down the plot and the importance of the \ion{Fe}{II} 4924 and \ion{Fe}{II} 5018 lines, and \ion{O}{I} and \ion{Ca}{II} NIR to the value of $N$ is also apparent.

\begin{table}
	\centering
	\caption{Common line blending combinations}
	\begin{tabular}{lccc}
	$N$ &  Description \\
	3 & \ion{Fe}{II} blend, \ion{Si}{II}, \ion{O}{I}/\ion{Ca}{II} blend \\
	4 & Same as $N=3$ but with visible \ion{Na}{I} \\
	6 & \ion{Fe}{II} 4924 and 5018 blended, all other lines separate \\
	7 & All lines separate\\    
	\hline
	\end{tabular}
	\label{tab:blends}
\end{table}

The blending of absorption features gives clues as to the shape of the density profile of the ejecta and, by proxy, \ek. In this section we discuss the situations in which lines transition from being blended to being separate.

\subsubsection{Line velocity and spread, and the effect on line blending}
The two main regions of interest are the blending of \ion{Fe}{II} 4924, \ion{Fe}{II} 5018, and the \ion{Fe}{II} 5169, 5198, 5235 group with each other and the blending of the \ion{O}{I} 7774 triplet and \ion{Ca}{II} NIR triplet. In velocity space the \ion{Fe}{II} lines are separated by $\sim 5600$ km/s (for \ion{Fe}{II} 4924 and 5018) and $\sim10,500$ km/s (for \ion{Fe}{II} 5018 and \ion{Fe}{II} 5169) while the \ion{O}{I} 7774 triplet and \ion{Ca}{II} NIR lines are separated by $\sim27,000$ km/s. This represents a good sampling of log-spaced velocity, which is why line blending is sequential. Thus from this cursory analysis we can expect that very high velocity ejecta is required to blend O and \ion{Ca}{II}, whereas the \ion{Fe}{II} lines require a less substantial velocity differential. Note that it is not sufficient for the ejecta to be high velocity; the values given here are the velocity differentials, that is the spread $\Delta v$ over which the atom or ion must provide sufficient opacity. If the lines were simply offset by $27,000$ km/s then \ion{O}{I} 7774 and \ion{Ca}{II} NIR would not blend, as can be seen in PTF12gzk \citep{Benami2012} where the line velocities are typically $>20,000$ km s$^{-1}$ but $\left<N\right>=6$. 

\subsubsection{Deblending} \label{sec:unblend}
\begin{figure*}
	\centering
	\includegraphics[scale=0.7]{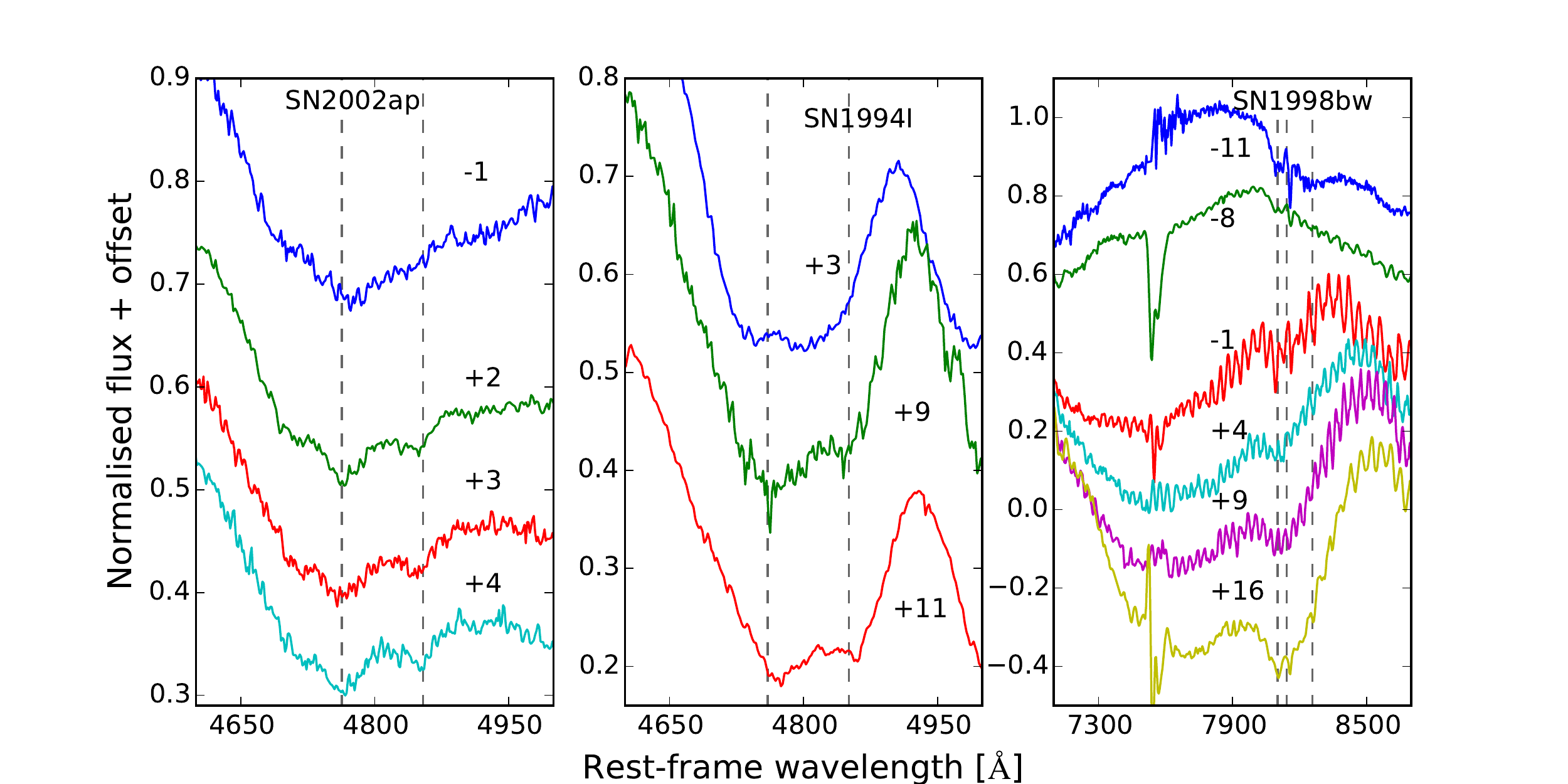}
	\caption{(Left) The evolution of the \ion{Fe}{II} 4924 and 5018 features in SN 2002ap; the grey dashed line represents the lines shifted by $\sim 10,000$ km s$^{-1}$. The two lines emerge out of the broader feature over a period of two days. (Centre) The same as SN 2002ap but for SN 1994I. The velocity of the \ion{Fe}{II} lines is $\sim 10,000$ km s$^{-1}$ but they are weaker and take longer to appear in terms of light curve evolution than for SN 2002ap. (Right) The evolution of the \ion{O}{I} and \ion{Ca}{II} blend for SN 1998bw. At $-$11 days there is an indication of material contributing to the opacity which would match \ion{Ca}{II} at $\sim15,000$ km s$^{-1}$, however it is not possible to say if this is \ion{Ca}{II} or not. After peak the \ion{Ca}{II} NIR triplet lines become prominent at a velocity of $\sim13,000$ km s$^{-1}$. }
	\label{fig:unblend}
\end{figure*}

Figure~\ref{fig:unblend} shows how various lines become deblended in the spectra, although it is noticeable that this tends to occur around or after peak in most SNe. The broad-lined supernovae provide a useful study into this process because they represent the largest possibility for change in the spectra. Lines deblend as the opacity velocity differentials decrease because of ejecta expansion and the changing slope of the density profile.

\subsection{$\left<N\right>$ as an aid to classification} \label{sec:class}
Having established how we count $N$ and what it represents we now consider $\left<N\right>$, the mean $N$, which is calculated from the $t_{-1/2}$ and $t_\mathrm{max}$ spectra and is used as a measure of the pre-peak spectral evolution. In the absence of spectra at $t_{-1/2}$ the earliest spectrum after this time is used, alternatively the value at $t_\mathrm{max}$ is used. If there are no spectra at $t_\mathrm{max}$ then $N(t_\mathrm{max})$ is interpolated from spectra before and after. $\left<N\right>$ is taken to be an integer and so non-integer values are rounded up. In general $N(t_{+1/2})$ cannot be used to reliably estimate $\left<N\right>$ and the issues affecting post-maximum classification of SE-SNe are discussed in Section~\ref{sec:pp}.

Using this information it is now possible to adjust the existing classification system to include information about the degree of line blending in the SNe, which is done by including $\left<N\right>$ in the ``Ic'' nomenclature and dispensing with the ambiguous ``Ic-BL''. For example, broad-line Ic SN 2002ap becomes ``Ic-4'', whereas narrow lined Ic SN 2007gr becomes ``Ic-7''. Table~\ref{tab:IcN} gives the new classification for the SNe used here. This method is advantageous because the relative degree of line broadening between SNe is clearly given. 
 
\begin{table}
	\centering
	\caption{Ic classification including information on $N$}
	\begin{tabular}{lc}
	SN & Ic$\left<N\right>$ \\
	1994I & Ic-6 \\
	1997ef & Ic-4 \\
	1998bw & Ic-3 \\
	2002ap & Ic-4 \\
	2003jd* & Ic-4 \\
	2004aw & Ic-6 \\

	2004dn & Ic-6 \\
	2004fe & Ic-6 \\
	2005az & Ic-7 \\
	2006aj$^\dag$ & Ic-6 \\
	2007gr & Ic-7 \\
	2009bb & Ic-6 \\
	2010ah & Ic-3 \\
	2010bh* & Ic-3 \\
	PTF10vgv & Ic-5 \\
	2011bm & Ic-7 \\
	2012ej & Ic-7 \\
	PTF12gzk & Ic-6 \\
	2016P & Ic-6 \\
	2016coi & Ic-4 \\
	2016iae & Ic-7 \\
	\multicolumn{2}{l}{$^\dag$ $\left<N\right>$ is an upper limit, from \tmax\ spectrum only}\\
	\multicolumn{2}{l}{*Lower limit due to contamination or poor S/N}
	\end{tabular}
	\label{tab:IcN}
\end{table} 

\begin{figure}
	\centering
	\includegraphics[scale=0.4]{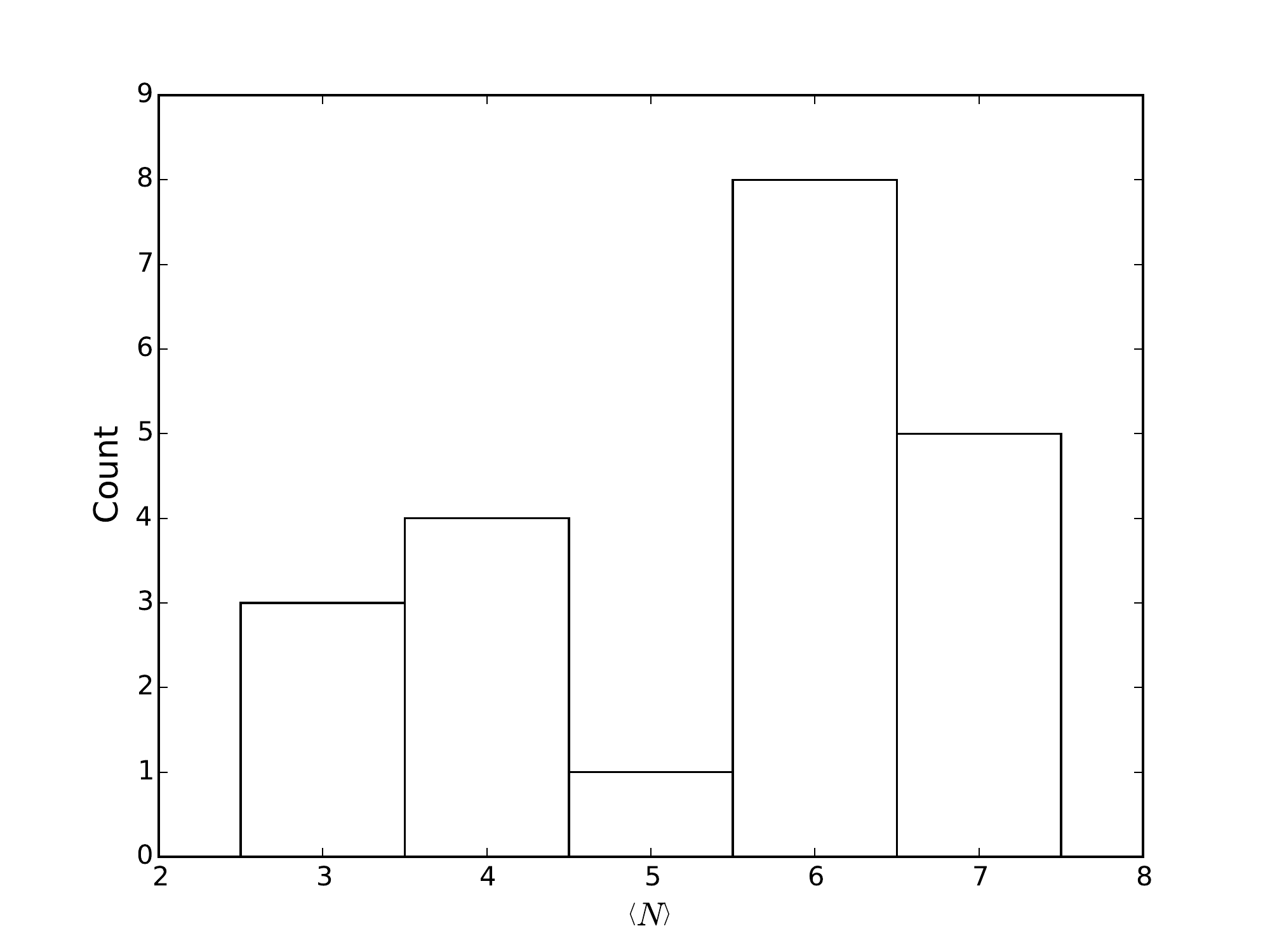}
	\caption{The number of SNe within the category Ic-$\left<N\right>$. The values $\left<N\right> = 3, 4, 6, 7$ represent particular sets of line blends. Those that fall in between are transitional between two groups. There is uncertainty in the categorisation for some SNe due to other sources of emission which is not reflected here.}
	\label{fig:nums}
\end{figure}

\begin{figure*}
	\centering
	\includegraphics[scale=0.7]{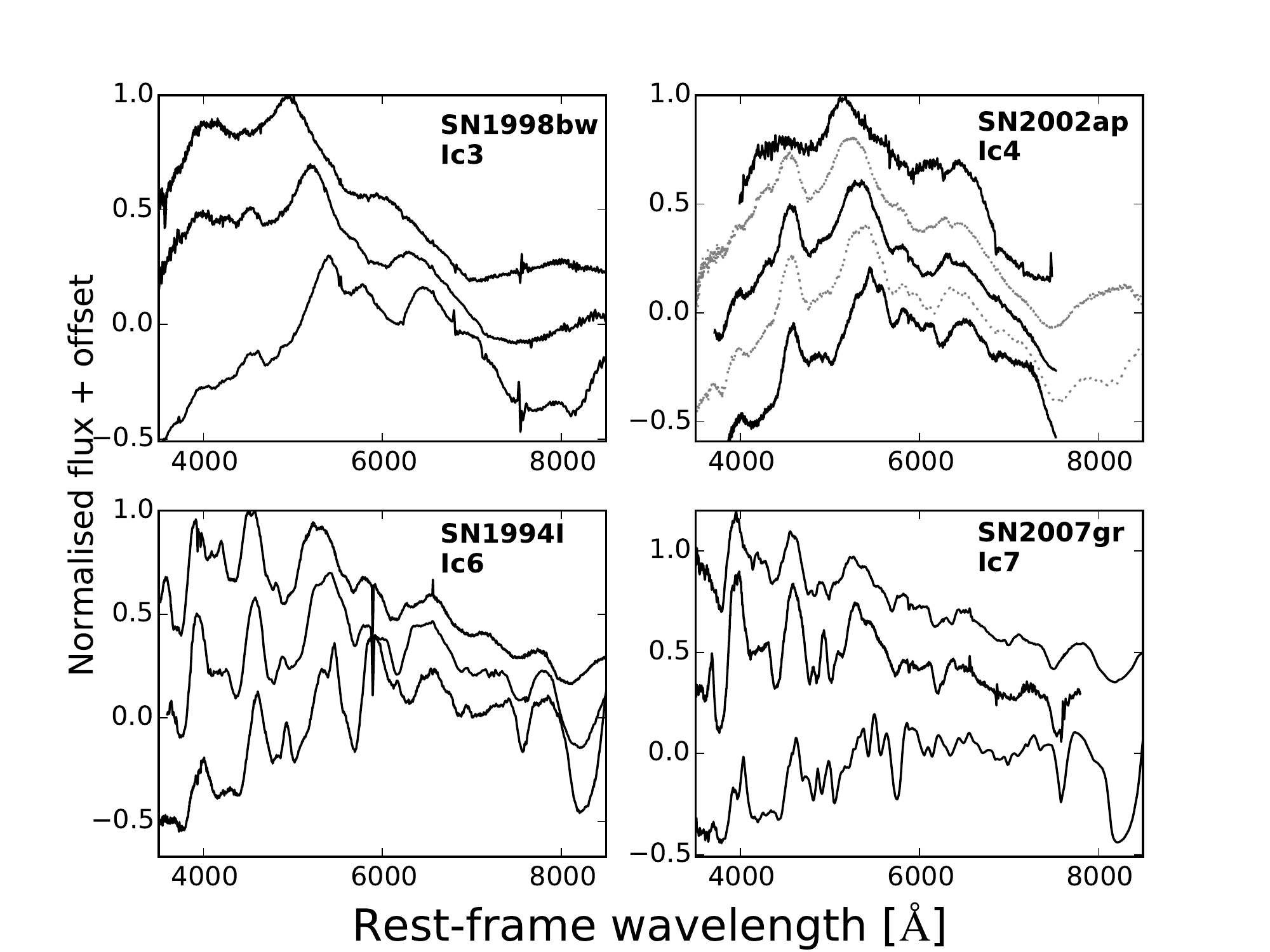}
	\caption{The $t_{-1/2}$, $t_\mathrm{max}$, and $t_{+1/2}$ spectra of four representative SNe for the SN Ic classification scheme presented in this work. Only the first two spectra are used in defining $\left<N\right>$, the last is given to show spectroscopic evolution. (Top left) GRB-SN 1998bw is a model for the Ic-3 group, and it displays blended \ion{Fe}{II} lines, \ion{O}{I} and \ion{Ca}{II}, and a single \ion{Si}{II} feature until well after maximum. The deblending of the \ion{Fe}{II} lines is affected by the depressed flux blueward of 5000 \AA . (Top right) Here SN 2002ap is used to show the effects of early deblending. The early spectrum shows similarity to SN 1998bw at a relative epoch. The grey dotted spectra are a few days later than the reference spectra and are used to show the evolution of the \ion{O}{I}/\ion{Ca}{II} blend. By maximum SN2002ap has deblended \ion{O}{I} and \ion{Ca}{II} lines, a distinct \ion{Na}{I} line and the \ion{Fe}{II} lines, while still blended, are beginning to separate. (Lower left) SN 1994I shows the evolution of Ic-6 SNe. Generally the only blend is of the \ion{Fe}{II} 4924 and 5018 lines and these separate shortly after maximum. Notice that, except for  similarities between SN 2002ap and SN 1994I at $t_{+1/2}$, particularly around 4500 - 5500 \AA\ and 6200 to 7000 \AA . (Lower right) Ic-7 are well demonstrated by SN 2007gr, the lines are narrow and well defined throughout.}
	\label{fig:Icclasses}
\end{figure*}

Figure~\ref{fig:Icclasses} shows representative spectra for SNe of type Ic-3, Ic-4, Ic-6, and Ic-7 while Figure~\ref{fig:nums} shows how many SNe fall into each Ic-$\left<N\right>$ category. That there appears to be as many SNe with significant line blending compared to those with well defined lines is partly a consequence of a bias in the prioritisation of SN observations and of these SNe tending to be more luminous on average  \citep{Prentice2016}.

\section{Taxonomy} \label{sec:8}
In Section~\ref{sec:class} $\left<N\right>$ was established as a means of characterising the spectra of SNe Ic, and in Appendix~\ref{sec:7} we compare $\left<N\right>$ with physical parameters derived from the bolometric light curves of the SNe. From this it can be seen that $\left<N\right>$ is weakly correlated with line velocity, as would be expected, but no correlation with any other property.
Consequently, with two addition parameters, we can further expand on this method by including $v_\mathrm{SiII}$ at peak in units of 1000 km s$^{-1}$, and $t_{+1/2}$ in days, which gives Ic-$\left<N\right>\left(v_\mathrm{p,SiII}/t_{+1/2}\right)$. 
These measurements provide a useful at-a-glance comparison between He-rich SNe.
This taxonomic scheme is applied to the He-poor SNe in our sample in Table~\ref{tab:Icclass}. Note that for GRB-SNe the reclassification does not supersede the ``GRB-SN'' label and we find that all SNe associated with GRBs are classified as Ic-3.

\begin{table}
	\centering
	\caption{The reclassification of SNe Ic }
	\begin{tabular}{lcc}
	SN & Previous classification & Reclassification \\
	1994I & Ic &Ic-6(11/9)\\
	1997ef &Ic-BL & Ic-4(13/45)\\
	1998bw & GRB-SN &Ic-3(15/16)\\
	2002ap & Ic-BL & Ic-4(9/16)\\
	2003jd* &Ic-BL & Ic-4(13/14)\\
	2004aw &Ic & Ic-6(11/21)\\
	2004dn &Ic &  Ic-6(9/15)\\
	2004fe & Ic & Ic-6(8/u)\\
	2005az & Ic &Ic-7(11/29)\\
	2006aj$^\dag$ & GRB-SN/Ic-BL & Ic-6(21/14)\\
	2007gr & Ic & Ic-7(7/15)\\
	2009bb & Ic-BL & Ic-6(20/13)\\
	2010ah & Ic-BL & Ic-3(18/17)\\
	2010bh* & GRB-SN & Ic-3(32/9)\\
	PTF10vgv & Ic & Ic-5(7/10)\\
	2011bm & Ic & Ic-7(6/43) \\
	2012ej & Ic & Ic-7(7/20)\\
	PTF12gzk & Ic & Ic-6(17/24) \\
	2016P & Ic-BL & Ic-6(u/14) \\
	2016coi & Ic-BL & Ic-4(14/21)\\
	2016iae & Ic & Ic-7(9/14) \\
	\multicolumn{3}{l}{*$\left<N\right>$ Lower limit due to contamination or poor S/N}\\
	\multicolumn{3}{l}{$^\dag\left<N\right>$ is estimated from \tmax\ only}\\	
	\multicolumn{3}{l}{u: insufficient data to calculate}\\
	\hline
	\end{tabular}
	\label{tab:Icclass}
\end{table}

\begin{figure}
	\centering
	\includegraphics[scale=0.43]{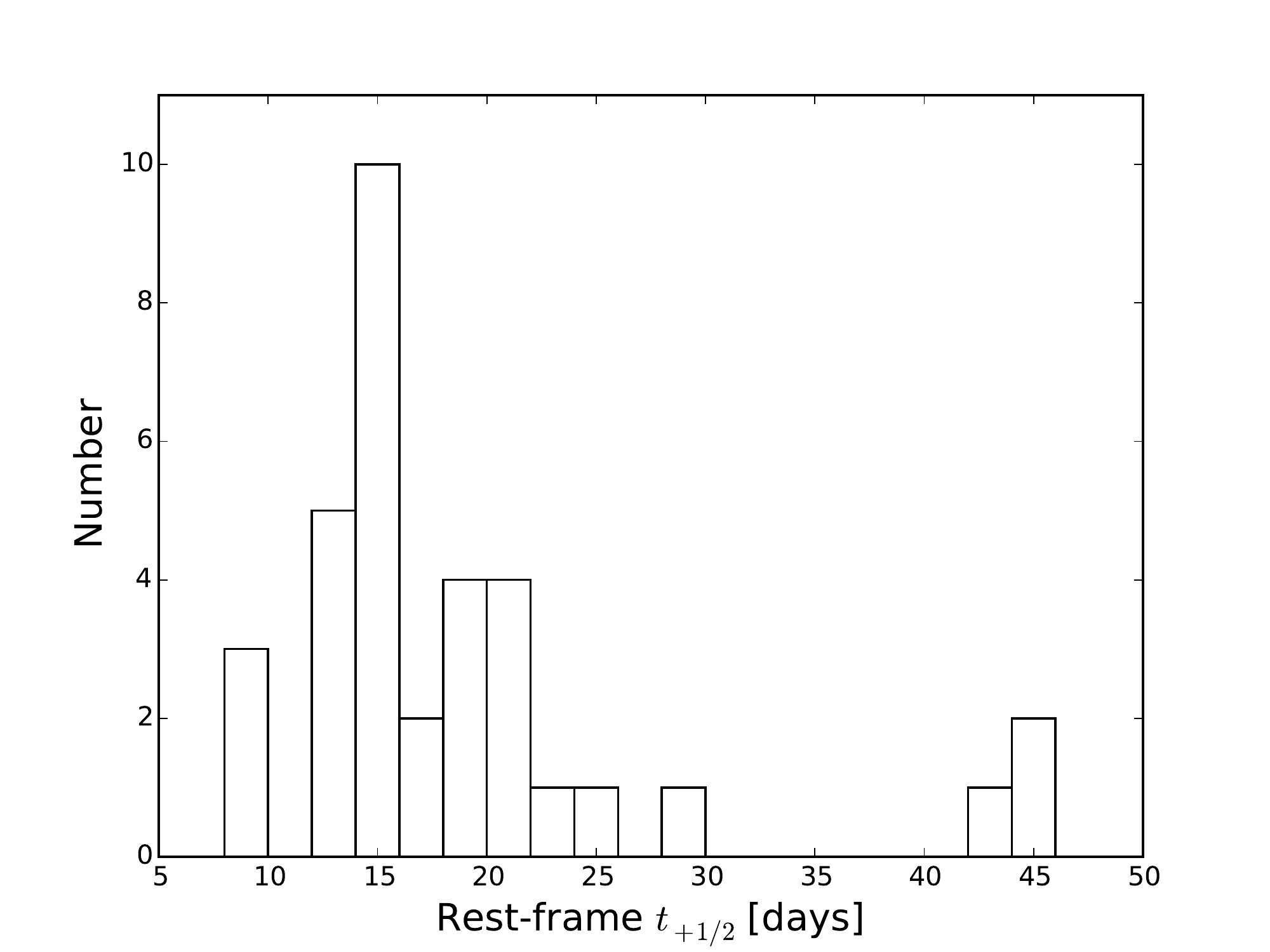}
	\caption{For reference, the distribution of $t_{+1/2}$ for 34 SNe Ic. As noted in Section~\ref{sec:N}, $t_{+1/2}$ is usually calculated from the bolometric light curve but the {\it VRgr} bands also provide good approximations.}
	\label{fig:risehist}
\end{figure}

Figure~\ref{fig:risehist} shows the distribution of $t_{+1/2}$ for all the SNe Ic in the sample used here, plus those from \cite{Prentice2016} that did not make the cut here. This figure is a useful reference when considering how a value of $t_{+1/2}$ sits in relation to other SNe. It is apparent that $t_{+1/2}<20$ d in most cases but that in a few extreme examples $t_{+1/2}$ can be around double this.

\section{Discussion} \label{sec:9}

\subsection{He-rich SNe}

The method for characterising the $\sim6200$ \AA\ region used for the He-rich SNe has allowed for the separation of these SNe into four separate groups defined by where they sit in the $\left<EW\right>$/$\left<f_\mathrm{em}/f_\mathrm{abs}\right>$ plane, which represents the degree of certainty that there is H in the ejecta. For the IIb and IIb(I) sub-groups this is definite as the H$\alpha$ P-Cygni profile is a  prominent feature and H$\beta$ and H$\gamma$ are present in most cases. 
For SNe Ib(I) the presence of H is likely, implied by the transition in strength of the $\sim6200$ \AA\ feature across the IIb(I) SNe to the Ib(II) SNe, but supporting evidence such as higher Balmer lines is absent and the spectra are dominated by strong \ion{He}{I} from early epochs. \ion{Si}{II} 6347 may well account for some of the opacity that results in the $\sim6200$ \AA\ absorption feature. The Ib group displays a weak absorption around the $6200$ \AA\ region, and an even weaker emission peak, that is most likely attributable to \ion{Si}{II} 6347, with little, if any, contribution from H$\alpha$. If \ion{H}{} is present in these SNe it is likely diffuse very soon after explosion and so provides little opacity.

The analysis presented here suggests that most He-rich SNe are not stripped down to their \ion{He}{} layer but retain some amount of \ion{H}{} in their ejecta.  Four of the SNe IIb (1993J  2011hs, 2011dh, 2011fu) display an early rapidly decaying phase in their light curve that can be interpreted as cooling of the stellar surface following shock-breakout. The timescale of shock-breakout, and its luminosity are related to the radius of the progenitor, ejecta mass, and explosion energy \cite[see][]{Nakar2014,Piro2015}. A more extended progenitor has a longer cooling timescale and greater luminosity, which is expected to be seen in the light curve as a rapid initial peak ($\sim$ half a day) followed by decline over a few days (for a more extended envelope). No such early time emission is seen for any other IIb or IIb(I) SNe in our sample and, while for most of the SNe this could be due to lack of detection at a sufficiently early time, the lack of a declining phase in the light curve puts an upper limit on the extent of the progenitor's envelope. This discrepancy has been used to make the case that SNe IIb can occur from both compact ($\sim$ a few solar radii) and extended ($\sim$ a few hundred solar radii) progenitors \citep[e.g.,][]{Bersten2012,Folatelli2014}.
Interestingly, two of the Ib(II) SNe have shown a similar emission; the peculiar SN 2008D/XRF080109 \citep{Mazzali2008,Soderberg2008,Modjaz2009}, and SN 1999ex \citep{Hamuy2002,Stritzinger2002} where the early emission has been attributed to shock-breakout from the stellar surface, or in the case of SN 2008D, to the failure of a relativistic jet to pierce the stellar envelope \cite{Mazzali2008}. Our results are consistent with more H-rich SNe also being more extended.

\subsection{He-poor SNe (SNe Ic)}
\subsubsection{A comparison of $E_\mathrm{k}$/$M_\mathrm{ej}$ with $N$}
It was noted in Section~\ref{sec:5} that a small amount of mass at high velocity can increase the kinetic energy of the ejecta by nearly 10$^{52}$ erg and that \eom\ was significantly affected by the shape of the outer density profile. The variation in density profile can be explained by small masses of material projected to high velocities, as could happen in the case of an asymmetric explosion. In this case we may expect to see widely separated peaks in the nebular phase \ion{[O}{I]} 6300, 6363 \AA\ line \cite[see][]{Mazzali2005,Maeda2008} and relatively broad, but not necessarily blended, lines for an off axis jet (e.g., SN 2003jd) or extremely broad and long-lived absorption features as with GRB-SNe if we view the jet on-axis (this would also explain why the GRB is detected in some Ic with low $\left<N\right>$ and not in others\footnote{Although the detection rate for GRBs is not 100 percent}). 

\begin{figure}
	\centering
	\includegraphics[scale=0.4]{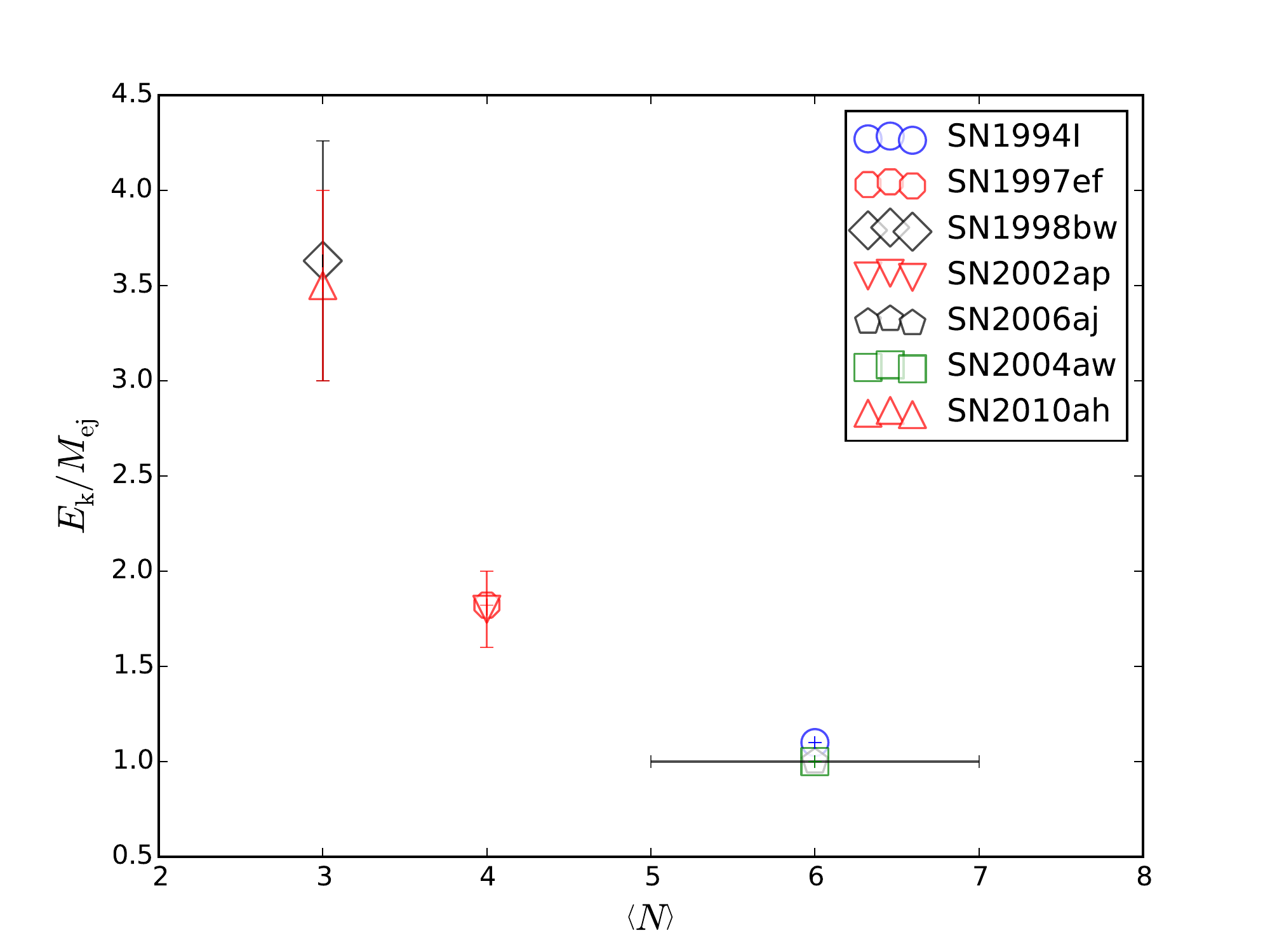}
	\caption{$E_\mathrm{k}/M_\mathrm{ej}$ as a function of $\left<N\right>$ for SNe with the former values derived from spectral modelling rather than analytical methods. The plot indicates that, for SNe with such values available, $\left<N\right>$ is related to $E_\mathrm{k}$/$M_\mathrm{ej}$. That the two parameters should be linked is a product of the outer density profile whereby a small amount of high velocity mass results in a large increase in kinetic energy. SN 2006aj is an unusual case because $\left<N\right>$ is difficult to define as it is likely that the spectra are contaminated by other sources of emission. This plot also demonstrates the need for more SNe Ic to be modelled spectroscopically as the commonly used analytical methods cannot account for the diverse range of density profiles.}
	\label{fig:EM}
\end{figure}

We compare $\left<N\right>$ with $E_\mathrm{k}/M_\mathrm{ej}$ derived from spectral modelling or hydrodynamic simulations (See Appendix~\ref{sec:ek} for the limitations of other methods). The number of SNe available for this kind of analysis is small, so we have just a few key examples which fortunately cover a good range of $E_\mathrm{k}/M_\mathrm{ej}$. Figure~\ref{fig:EM} shows the relationship between $E_\mathrm{k}$/$M_\mathrm{ej}$ and the number of features for the SNe that fulfil our requirements. It can be seen that $E_\mathrm{k}/M_\mathrm{ej}$ scales with smaller $\left<N\right>$ although the case of SN 2006aj is extremely uncertain as the pre-max spectra show contamination from the host-galaxy and the XRF afterglow \citep{Pian2006}.

The density profile of the ejecta is an important part of what defines the value of $E_\mathrm{k}/M_\mathrm{ej}$ \citep{Mazzali2013}. It is typically modelled with a steep index before reaching a turn-over at lower radial coordinates \citep{Mazzali2000}.
In a simple model we can define three cases:

Case 1: A steeper density profile provides more mass towards the core, at lower velocities. The photosphere evolves slower through velocity space in this environment as the increasing density of the material passing through the photospheric boundary offsets the decrease in density due to expansion. Line formation should be relatively narrow in velocity space as the opacity outside the photosphere rapidly drops due to the steep profile. This is typical of larger $\left<N\right>$.

Case 2: Flatter density profiles can project more material to higher velocities. For a fixed mass $M$ such a profile would have a higher overall $E_\mathrm{k}$ than for a steeper density profile, however the photosphere would recede more rapidly in velocity space. Such a profile should allow somewhat broad, but not necessarily blended, lines which would reduce in velocity rapidly. $E_\mathrm{k}/M_\mathrm{ej}$ would be larger than for Case 1.

Case 3: A steep density profile with a flatter outer part has already been discussed, but to summarise it would result in long-lived blended lines as the photosphere recedes slowly in the steep part of the density profile but the flatter part provides sufficient opacity to greater velocities resulting in line blending. This allows the larger mass of the steeper density profile to combine with the larger kinetic energy of the flatter density profile leading to a larger $E_\mathrm{k}/M_\mathrm{ej}$ than Case 2. This situation would result in a very low $\left<N\right>$.

A relationship between $\left<N\right>$ and the outer density index $n$ is suggested in that a lower value of one represents a lower value of the other. For SN 1998bw $n=2$ and $\left<N\right>=3$, while for SN 1994I $n=7$ and $\left<N\right>=6$. It may be possible that with more work a set of models could be used to provide an analytical estimate of the total kinetic energy of the SN by using $\left<N\right>$ as a proxy for $n$.

\subsubsection{Post-peak spectral similarity} \label{sec:pp}

\begin{figure}
	\centering
	\includegraphics[scale=0.43]{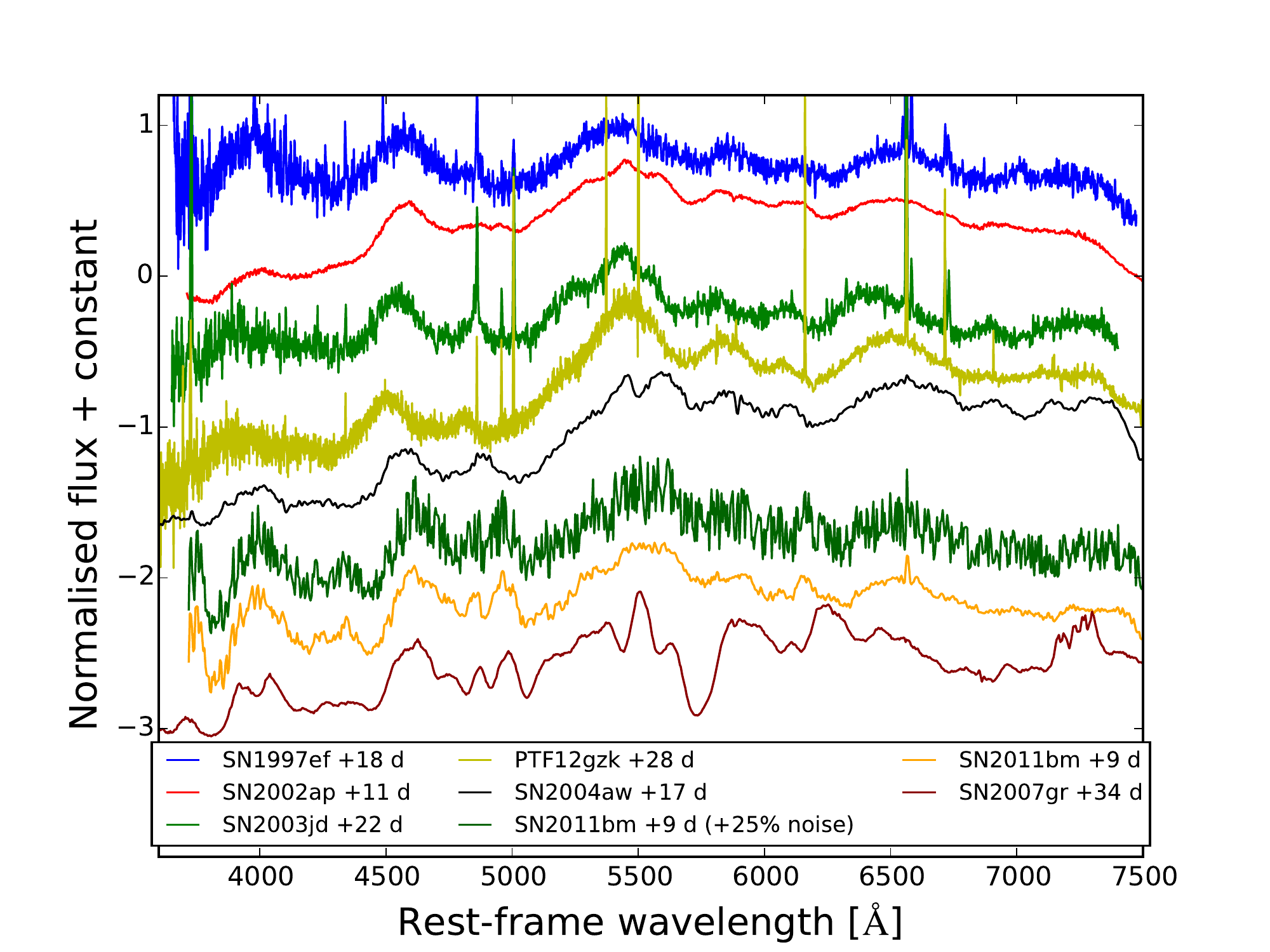}
	\caption{Various spectra that appear somewhat similar post-maximum, this is particularly the case for the top five SNe for which $\left<N\right>=3,3,3,5,5$ respectively . Only the spectra of Ic-7 SN 2011bm and Ic-7 SN 2007gr show clear deviation from the general characteristics of the other spectra. We include a version of the observed SN 2011bm spectrum where 20 percent noise has been introduced, which shows that S/N is important for classification after peak as this spectrum could be mistaken for belonging to the group of spectra above it. }
	\label{fig:late}
\end{figure}

We have considered spectra at times before and around maximum light. However, many SNe are classified at a later epoch and while it is clear that early spectra, even with relatively poor S/N, can be categorised but it is not clear that this true of later times. Table~\ref{tab:peakdecay} indicates that the line velocities of many SNe fall within a relatively narrow range that does not correspond with their velocities at earlier times. In Figure~\ref{fig:late} we plot the spectra of several supernovae where the spectra appear to be similar. The epochs are all post-maximum but vary in relation to the evolution of the light curve (i.e., they are not all around $t_{+1/2}$). 
While the order of spectra are grouped in terms of early spectral similarity it is clear that, aside from the Ic-7 SNe (SN2011bm and SN 2007gr, where the narrow lines remain prominent), the remaining SNe have spectral similarity on at lease one occasion during their evolution. To test the effect of noise on the Ic-7 SNe we include two versions of the same spectrum of SN 2011bm in which we artificially introduce noise at the 20 percent level to one of the spectra. Both the noisy and observed spectra show similarity to the spectra above them in the region redward of 5000 \AA. The key difference is in the region between $4000 - 5000$ \AA\ where the \ion{Fe}{II} lines lose their definition with increasing noise. While it is still possible to see a positive comparison between the two SN 2011bm spectra and that of SN 2007gr it is also possible to see similarity between the noisier SN 2011bm spectrum and the spectra above. A further increase in noise also increases the ambiguity of the spectrum and could make misclassification more likely. Such problems may well occur often as classification spectra are typically taken with shorter exposure times than science spectra and consequently a lower S/N. This can have an effect on the statistics of SN rates as post-peak classification is ambiguous.

\subsubsection{Sample size}
Our sample has relatively few SNe with spectra or light curves early enough to investigate $N$ at $t_{-1/2}$. Although this number increases for $t_\mathrm{max}$, it is evident that with 5 subgroups more data is required in order to effectively probe the parameter space. We can trace three issues that result in a dearth of numbers. The first is biases in observations, where rare SNe such as GRB-SNe or those with broad lines are followed in more detail than normal SNe Ic. The second is data with low S/N; it is understandable that spectroscopy can be time intensive for all but the brightest targets but information on the SN is lost if S/N is too low. The third is the lack of data made publicly available, which is especially relevant to spectra which cannot be accommodated in the literature in the way light curve information can. A significant proportion of the data on WiseRep has been supplied by 2014 Harvard-Smithsonian Centre for Astrophysics (CfA) data release \citep{Modjaz2014}. More data will allow more robust analysis to build on this work.

\section{Conclusions} \label{sec:10}

The publicly available spectra for the SNe in the sample of \cite{Prentice2016}, plus that of SN 1997ef, SN 2004dn, and unpublished data for SN 2012ej, SN 2016P, SN 2016coi, and SN 2016iae has been analysed, using an empirical method to group the spectra according to the presence and strength of \ion{H}{} lines, \ion{He}{} lines, and line broadening. From this, there was found to be a clear distinction between SNe of Type Ic (He-poor) and those of Type Ib/IIb (He-rich), the analysis then proceeded along two separate pathways

\subsection{He-Rich SNe}
The He-rich SNe were analysed for the presence and strength of \ion{H}{} in their spectra. Measurements were made of the velocity, equivalent width, and ratio of absorption to emission of the feature around $6200$ \AA\ which is typically attributable to H$\alpha$. An alternate explanation for this feature could be \ion{Si}{II} 6347 as is found in Type Ia and Ic SNe, however the analysis presented here suggests that, while \ion{Si}{II} 6347 may contribute to the opacity in this region, H$\alpha$ is more likely the dominant component for most He-rich SNe. 

The spectra were also examined for the presence of Balmer lines beyond H$\alpha$, which are prominent in H-rich SNe but rapidly decrease in strength with decreasing H$\alpha$ strength. It is found that some classical IIb SNe do not display clear H absorption beyond H$\alpha$ while some SNe Ib show hints of these features that are broadly in line with a continuum of \ion{H}{} line strength. 

The mean pre-peak contrast ratio between $\left<f_\mathrm{abs}/f_\mathrm{em}\right>$ and the mean pre-peak equivalent width $\left<EW\right>$ have been used to systematically categorise the $\sim6200$ \AA\ feature, and from this the He-rich SNe are able to be placed into four groups:
\begin{itemize}
	\item{IIb - H-rich, H$\alpha$ emission dominates absorption}
	\item{IIb(I) - moderately H-rich, H$\alpha$ absorption dominates emission}
	\item{Ib(II) - Likely shows some H$\alpha$ but lacks any definite signatures. Like IIb(I) but with a weaker line profile}
	\item{Ib - Weak 6200 \AA\ feature, probably \ion{Si}{II} dominated}
\end{itemize}

An emission dominated H$\alpha$ feature relates to a strong P-Cygni line profile, which is indicative of an extended H envelope. Such a large distribution of emitting material in the outer ejecta is incompatible with the location and abundances of Si found in He-rich SNe. Finally, the results here are consistent with the conclusions driven from analysis of the early time declining phase of the light curve of some SNe IIb in that the most H-rich SNe are also the most extended because their outer envelope would be H dominated.

\subsection{Type Ic SNe}
The method adopted to analyse the spectra of type Ic SNe was to count the number of absorption features $N$ present in the spectra at key epochs. 7 key features have been selected due to their prominence and ubiquity in SE-SNe spectra, because they allow for an investigation into the velocity spread over which blending occurs, and due to the wavelength constraints of commonly used spectrographs. Because line blending is related to the specific kinetic energy of the ejecta this is a useful diagnostic for classification and provides information about the properties of the SNe without resorting to more complex processing methods. 

By taking a mean value for $N$ before bolometric maximum and comparing this value with the velocities of \ion{Si}{II}, \ion{Fe}{II}, and \ion{O}{I} it is found that there is a relationship between $\left<N\right>$ and line velocity, although this should be expected as sufficient mass at high velocity, along with a steep density profile in the outer ejecta, is required to form broad lines. Comparison with the properties of the SNe light curves ($L_\mathrm{p}$, $t_{-1/2}$, \tdecay) as well as $M_\mathrm{Ni}$ reveals that there is no connection between $\left<N\right>$ and any of these properties. However, the specific kinetic energy $E_\mathrm{k}$/$M_\mathrm{ej}$, using values derived from spectral modelling or hydrodynamical simulations, may be related to $\left<N\right>$. This is because line blending requires an outer, flatter component to the density profile of the ejecta which increases the mass at high velocity, and with it $E_\mathrm{k}$, but without significantly increasing the overall mass. This could be a consequence of ejecta with a jet like structure as is expected with GRB-SNe. High velocities alone do not translate into a large $E_\mathrm{k}$/$M_\mathrm{ej}$. If the lines are relatively narrow then the density profile is steeper in front of the photosphere. 

To reflect the findings here we adapt the common nomenclature for SN classification by including $\left<N\right>$, $v_\mathrm{SiII}$ at peak, and $t_{+1/2}$ so as to give a clearer picture of the degree of line blending in SNe Ic and to provide physical information about the explosion. The adopted form of the taxonomical system is as follows: Ic-$\left<N\right>\left(v_\mathrm{p,SiII}/t_{+1/2}\right)$, with $v_\mathrm{SiII}$ in units of 1000 km s$^{-1}$. With this modification the properties of SNe Ic can be compared at a glance and the arbitrary nature of the ``broad-line'' definition is removed.

\subsection{Future work}
Finally, we have identified several SNe in our sample which could be subject to analysis through spectral modelling of early time spectra. $M_\mathrm{ej}$ and $E_\mathrm{k}$ are important physical parameters that relate to the explosion mechanism and central engine and require more detailed analysis. The diversity of explosion properties amongst SE-SNe means that more data from unbiased sources are necessary to examine the properties of the population as a whole. From this, the range of potential explosion mechanisms and energies, connection to high energy transients, progenitor mass distributions, estimates of binary fractions, and the potential for aspherical explosions (which in turn impacts the use of SE-SNe as electromagnetic counterparts to gravitational waves) can be explored.

%



\bibliographystyle{mnras}
\bibliography{specbib} 



%
\appendix
\section*{Acknowledgements}
We thank Phil James and Chris Ashall for useful comments. This research was funded by an STFC grant.
\section{The argument for H in He-rich SNe}\label{sec:app}

\subsection{H$\alpha$ or Si II 6347?} \label{sec:si}

\begin{figure}
	\centering
	\includegraphics[scale=0.4]{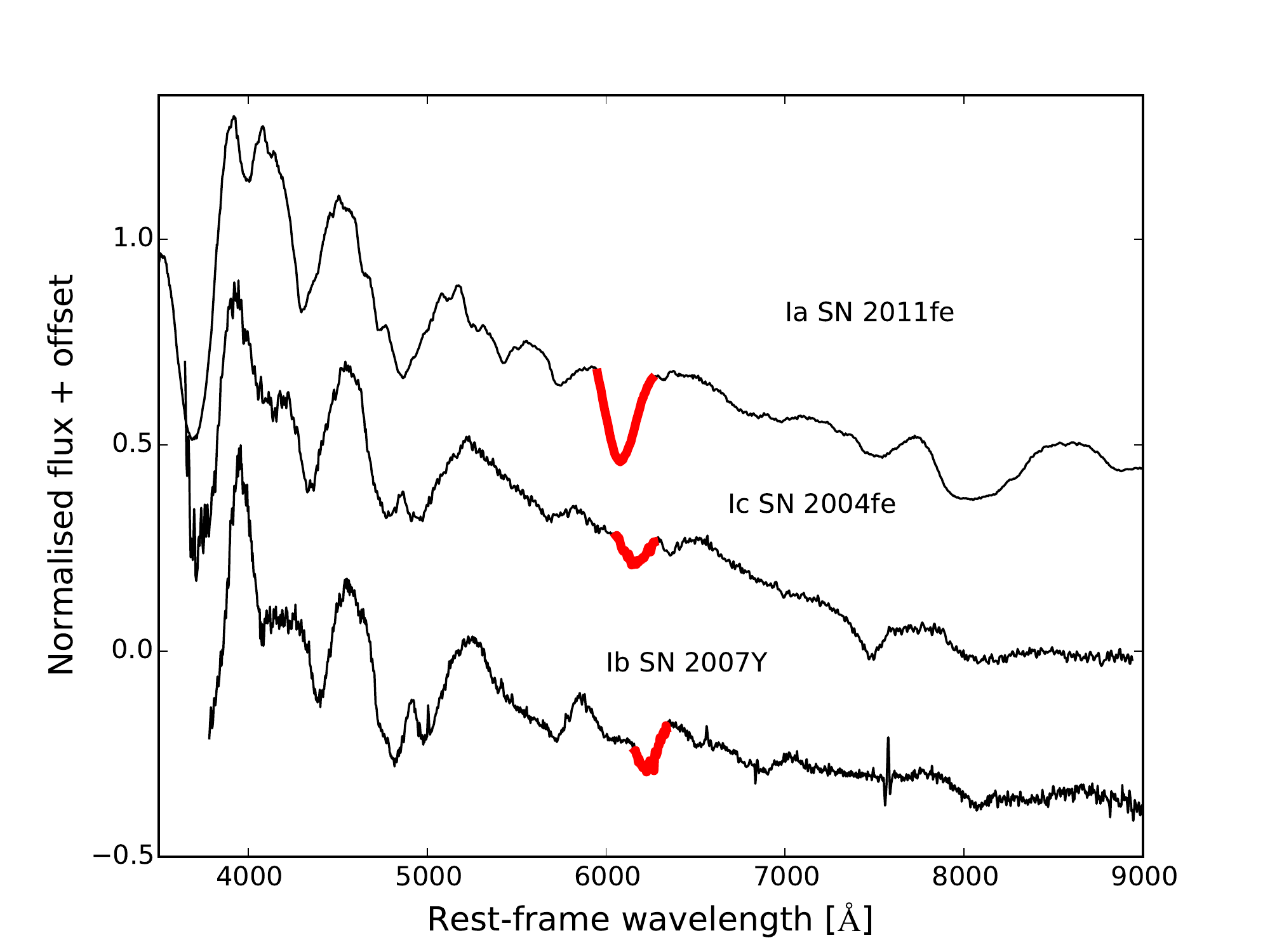}
	\caption{A comparison of the $\sim 6200$ \AA\ feature between SN Ia SN 2011fe \citep{Mazzali2014}, SN Ic 2004fe \citep{Modjaz2014}, and SN Ib 2007Y \citep{Stritzinger2009}. In SNe Ia and Ic this feature is attributed to \ion{Si}{II} 6347, for SNe Ib such an assertion is less certain due to the presence of \ion{He}{} in the outer ejecta which means that \ion{Si}{} is buried deeper in the ejecta and there is a possibility of H in the outer layers of the ejecta. In SNe Ic and Ia the feature immediately redward of \ion{Si}{II} 6347 is \ion{C}{II} 6580 while in SNe Ib this line is dominated by \ion{He}{II} 6678.}
	\label{fig:Sidemo}
\end{figure}

The positive identification of H$\alpha$ in H-poor SNe is complicated due to the possible presence of the \ion{Si}{II} 6347 \AA\ line. This line is such that it differs from H$\alpha$ by $\sim 10,000$ km s$^{-1}$, meaning that the $\sim 6200$ \AA\ feature could represent \ion{Si}{II} at  $6,000$ km s$^{-1}$ or H$\alpha$ at $16,000$ km s$^{-1}$, both of which are valid velocities for these atomic species in SN ejecta. The presence of \ion{Si}{II} in SE-SNe is not unexpected because SNe Ic, like SNe Ia, show an absorption feature associated with this ion \citep[see, for example,][]{Parrent2016}. We show the $\sim 6200$ \AA\ region for examples of SNe Ib, Ic, and Ia in Figure~\ref{fig:Sidemo}.

In SNe IIb the absorption component of the H$\alpha$ P-Cygni profile dominates any contribution from \ion{Si}{II}, which is also likely constrained to a shell that is below the photosphere for some significant period of the evolution of the SN. Indeed, the unambiguous identification of this ion in SNe IIb is extremely difficult as the H$\alpha$ line remains sufficiently dominant in the spectrum well past the photospheric phase ($\sim+40$ days). 

If we consider SNe Ib, the lack of higher Balmer lines in the spectra, in addition to the potential presence of \ion{Si}{II}, makes identification of the atomic species responsible for the feature at $\sim 6200$ \AA\ ambiguous. Matters are further complicated by the evolution in the spectra. We would expect H lines to be stronger at very early times because little else is above the photosphere and \ion{Si}{II} should be restricted to layers well below that of the H envelope \citep{Sauer2006,Nakamura2001,Hachinger2012}. As time progresses a \ion{Si}{II} line should become more prominent as the photosphere recedes in velocity space, revealing the deeper laying ejecta. Conversely, H should reach a limiting velocity corresponding to the bottom of the shell with the line becoming progressively weaker as the ejecta expand and the optical depth of the H envelope decreases. If the relative velocity of the two shells is $\sim 10,000$ km s$^{-1}$ then it may be possible for the $\sim 6200$ \AA\ feature to transition from being H-dominated to being \ion{Si}{II}-dominated, with little more than an asymmetrical absorption component to indicate that this has happened.

\begin{figure}
	\centering
	\includegraphics[scale=0.4]{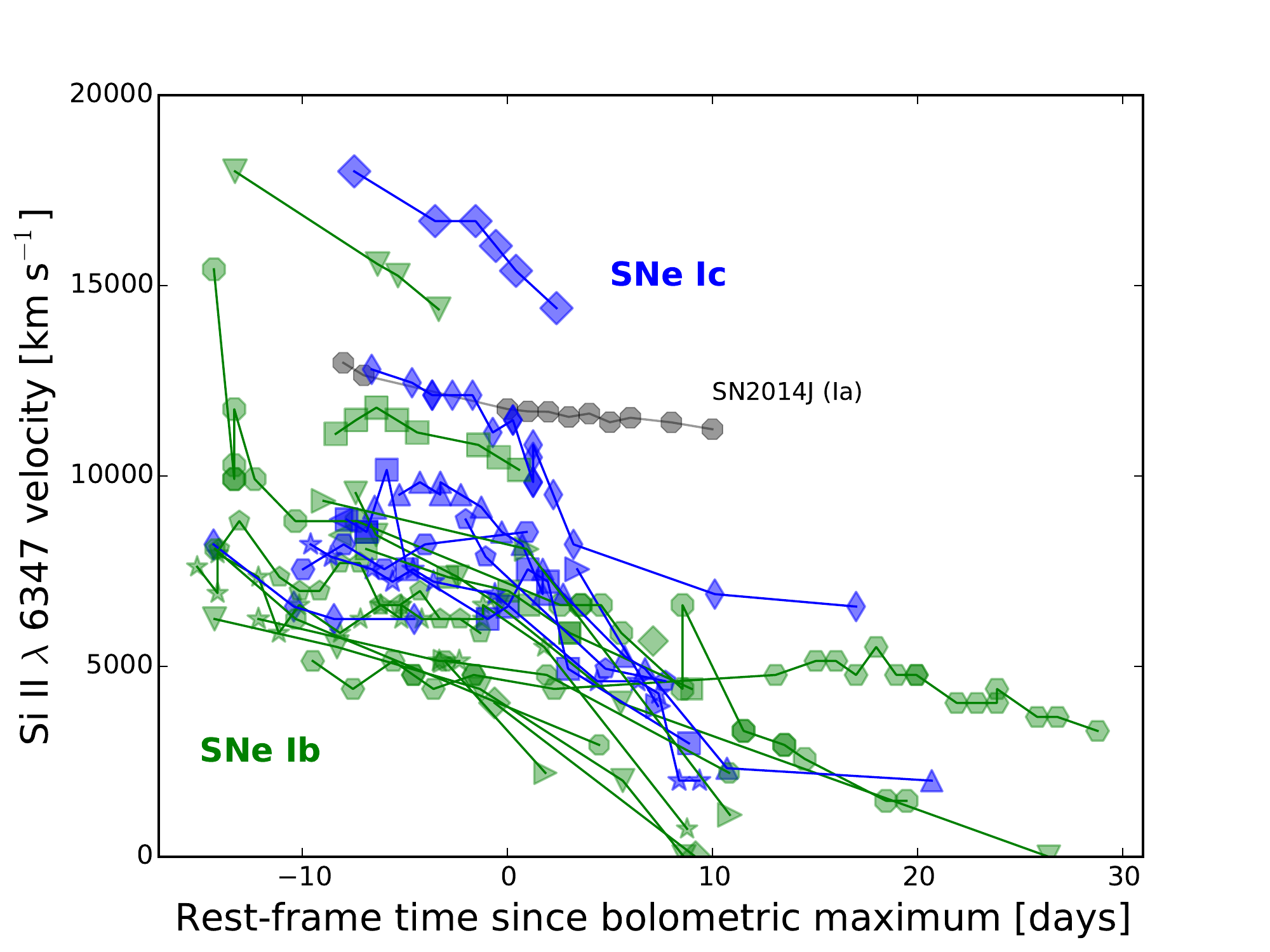}
	\caption{The velocities of the absorption line $\sim6200$ \AA\ if attributed to \ion{Si}{II} $6347$. The line is followed from the time it unambiguously appears until it disappears or the feature is ambiguous. On average the velocity for SNe Ib is lower than that for the Ic SNe, though there are exceptions, notably Type Ib SN 2007uy where the velocity is extremely large. For some of the SNe Ib the feature drops below the rest-frame wavelength of \ion{Si}{II} 6347, an unphysical situation indicating that the line is not produced by Si. For comparison we also give the \ion{Si}{II} velocity for Type Ia SN 2014J \citep{Ashall2014}}
	\label{fig:siiivel}
\end{figure}

We investigate this degeneracy by calculating the velocity of the $\sim 6200$ \AA\ feature for our SN Ib sample and for narrow-lined SNe Ic while insisting on a logical stratification to the ejecta \citep{Hachinger2012,Iwamoto1994} so that $v_{\mathrm{H}} > v_{\mathrm{He}} > v_{\mathrm{Si II}}$. The final discriminator is to ensure that the absorption feature does not extend significantly redwards of 6347 \AA , the rest wavelength of the \ion{Si}{II} line, as this would represent an unphysical situation. The result is shown in Figure~\ref{fig:siiivel}.

With the exception of SN 2007uy and PTF12gzk most of the SNe occupy a similar region in the time/velocity plane. However, the velocities of SNe Ic are typically higher than those of SNe Ib. In addition to this, some SNe Ib show velocities that drop below that of the rest wavelength of the \ion{Si}{II} line, which is a situation that is irreconcilable with the line being of that element. In most cases, for both SN types, the line disappears before it reaches a static velocity that would be indicative of the base of a shell. 

The disappearance of the line in SNe Ib represents a serious problem if we are to attribute the feature at $\sim6200$ \AA\ to \ion{Si}{II}. With a He shell above the Si layer, the \ion{Si}{II} strength should increase with time as more Si is exposed by the receding photosphere. Instead we see that the $\sim 6200$ \AA\ feature is strong initially (line strength for this feature is presented via equivalent width in Section~\ref{sec:ew}) before fading and disappearing around ten days after peak. This is likely an issue with opacity, where the line forming region has insufficient density and so $\tau\sim1$. If this is the case then the line-forming region should be in the outer layers of the ejecta, which is inconsistent with the position of a Si shell. For SNe Ic there is no He envelope and so Si should be exposed rapidly as the photosphere sits within the CO core material.

This reasoning does not mean that the presence of the \ion{Si}{II} ion can be ruled out, but it does mean that H is taken to be a valid source for the line opacity and so we henceforth refer to the $\sim 6200$ \AA\ feature in SNe Ib as H$\alpha$.

\subsection{Is there evidence of H$\alpha$ in the nebular phase?}
\subsubsection{Absorption}
\begin{figure}
	\centering
	\includegraphics[scale=0.4]{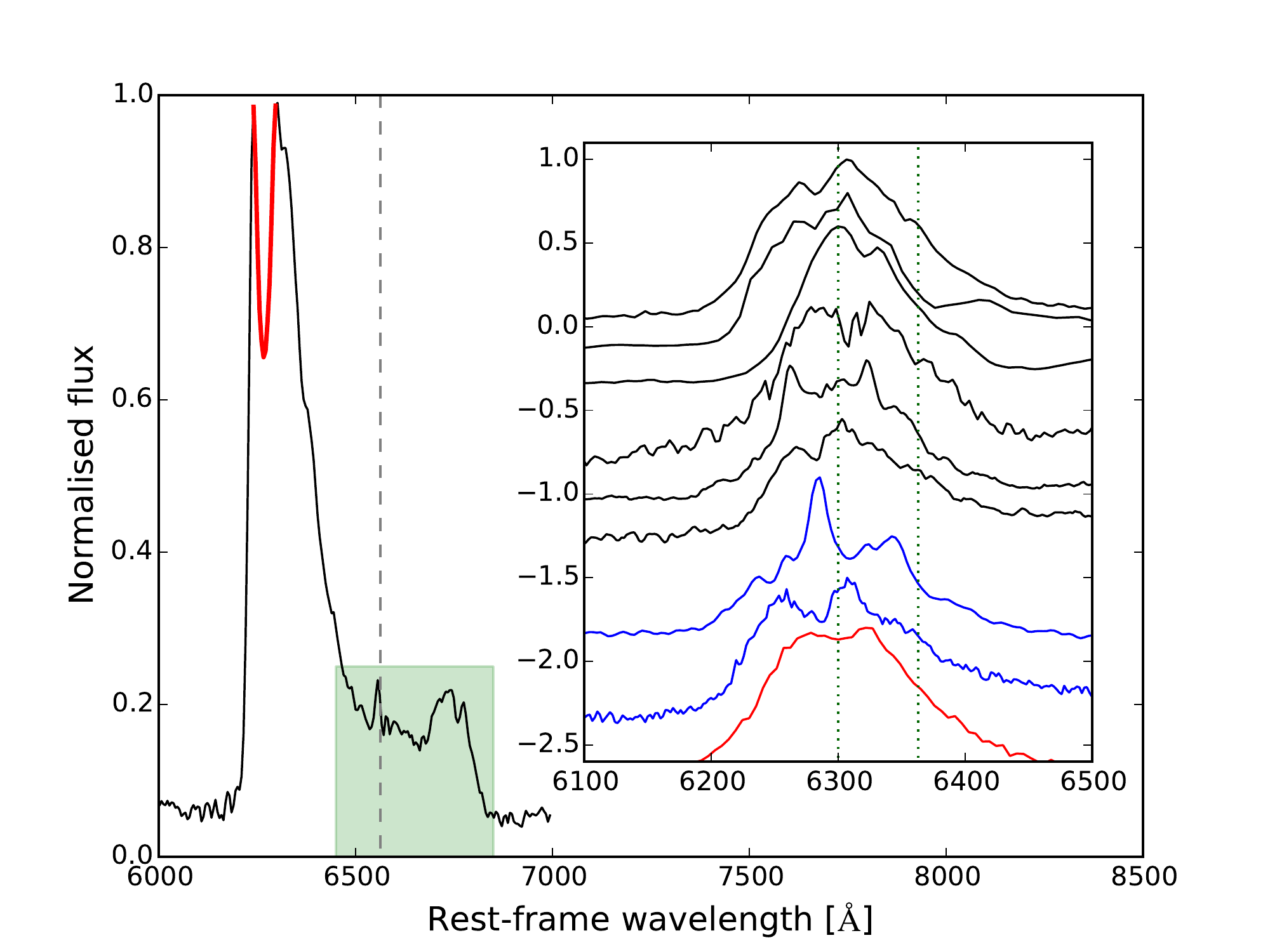}
	\caption{The nebular profile of Type IIb SN 2011fu at $+281$ days. The green region highlights the ``flat-top'' emission profile that is occasionally seen in He-rich SE-SNe, the dashed line at $6563$ \AA\ gives the rest wavelength of H$\alpha$, and the red region shows a possible absorption by H$\alpha$ on the \ion{[O}{I]} 6300, 6363 emission line. (Inset) The two dotted lines at 6300 and 6363 \AA\ demonstrate that although the \ion{[O}{I]} is a doublet, of which the 6300 \AA\ component is the stronger part of the two, the split feature is not due to the doublet nature of the line. This line profile is common, but not ubiquitous, to He-rich SE-SNe. Various SNe IIb are in black, SNe Ib 2009jf and 2008D in blue, and Type Ic SN 2011 is included for reference in red.}
	\label{fig:flattop}
\end{figure}

\begin{figure}
	\centering
	\includegraphics[scale=0.4]{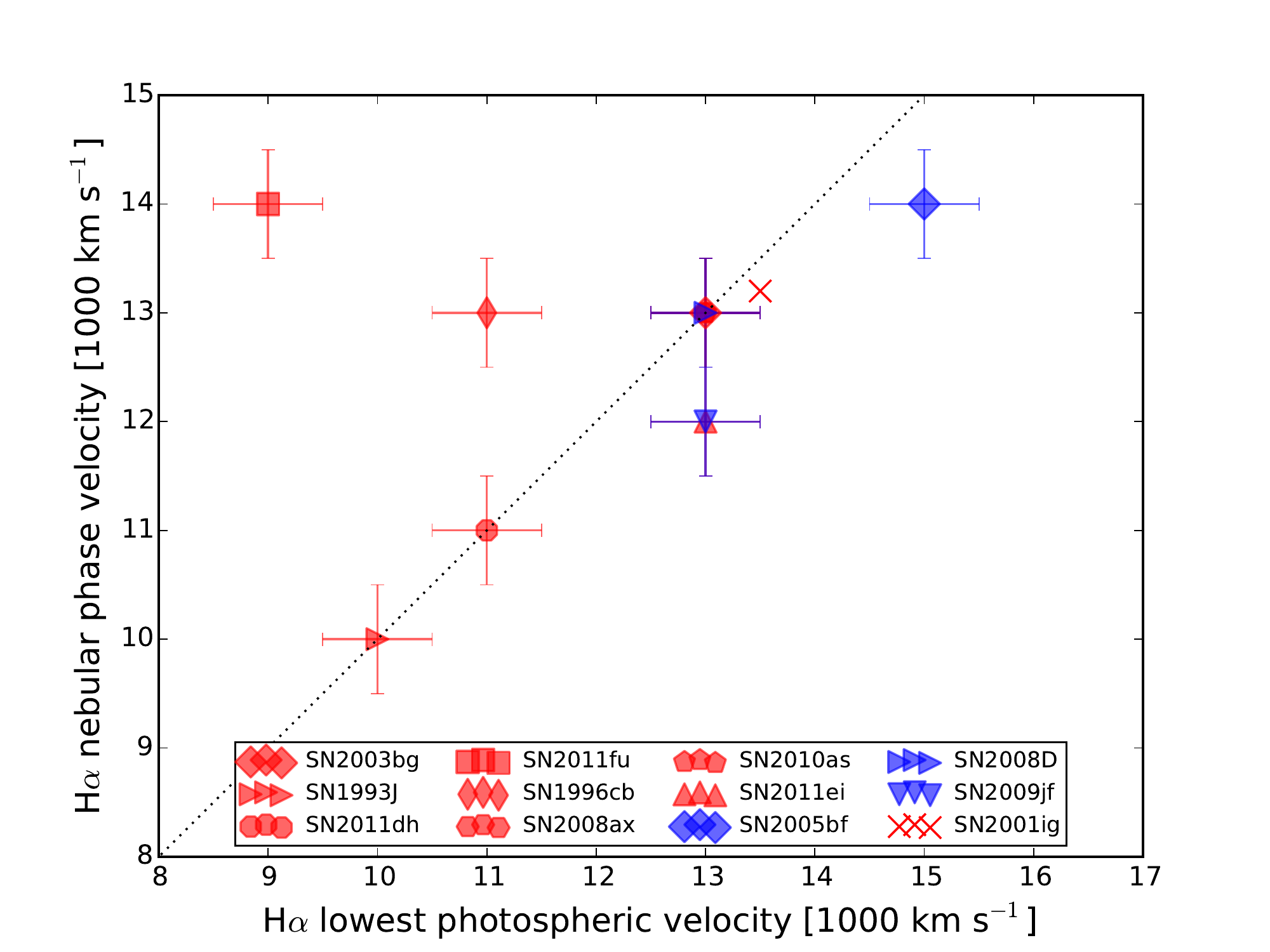}
	\caption{The velocity of the H$\alpha$ absorption profile on \ion{[O}{I]} 6300, 6363 during the nebular phase as a function of minimum photospheric velocity. for the supernovae in our sample and that of SN 2001ig \citep{Maurer2010}. Velocity measurements were approximated to the nearest integer value and an uncertainty of $\pm0.5$ included. There is a trend in some of the SNe for the two velocities to take similar values but there are also some clear outliers, most notably SN 2011fu where $v_\mathrm{neb}$ is amongst the largest measured and $v_\mathrm{phot}$ the lowest.  }
	\label{fig:Hacomp}
\end{figure}

Modelling of the nebular data of SN 1993J at $\sim 150$ days by \cite{HF1996} indicates that an absorption feature around $6380$ \AA , on the blue-wing of the \ion{[O}{I]} 6300, 6363 \AA\ line, is H$\alpha$ absorption corresponding to a shell with width $\sim 1000$ km s$^{-1}$ and an expansion velocity of $\sim 10,000$ km s$^{-1}$. \cite{Maurer2010} find that similar absorption features can be found in the \ion{[O}{I]} line for other SNe of Type Ib and IIb, as demonstrated for SN 2011fu in Figure~\ref{fig:flattop}, which they attribute to H$\alpha$. Thus, from this absorption feature a velocity can be calculated which would be expected to correspond to the densest part of the H shell, the base. The nebular H$\alpha$ velocity as a function of the minimum photospheric H$\alpha$ velocity is shown in Figure~\ref{fig:Hacomp}.

Another cause of a double-peaked \ion{[O}{I]} emission line is asphericity \citep{Mazzali2005,Maeda2008}, whereby the principle mass of \ion{O}{} is ejected in a torus leading to a line profile that is viewing-angle dependent. This could be typical of a more energetic explosion, as is characteristic of some SNe Ibc, while the only comparatively energetic Type IIb is SN 2003bg \cite{Mazzali2009,Hamuy2009}. It cannot be discounted that there is some degree of asphericity in the explosion of SNe IIb contribute to the splitting of the \ion{[O}{II]} line. Indeed, both SN 2005bf \citep{Tominaga2005} and SN 2008D \citep{Mazzali2008} have been suggested to have significant deviations from spherical symmetry and their nebular \ion{[O}{I]} line profiles are split. Another clue can be determined by the profile of \ion{[O}{I]} 5577, where by intrinsic asymmetries should be apparent in this line as well, but this is not seen.

\begin{table}
	\centering
	\caption{The profile of \ion{[O}{I]} at nebular epochs }
	\begin{tabular}{cccc}
	SN & Type & SN & Type\\
	\multicolumn{2}{c}{Double} & \multicolumn{2}{c}{Single}\\
	1993J & IIb & 2007Y & Ib\\
	1996cb & IIb & 2011hs* & IIb \\
	2003bg & IIb & 1999dn & Ib \\
	2008ax & IIb & 2007kj & Ib \\
	2011dh & IIb & iPTF13bvn & Ib \\
	2011ei & IIb & 2005hg & Ib \\
	2011fu & IIb & 2007C & Ib \\
	2005bf & Ib & & \\
	2010as & IIb & & \\
	2008D & Ib & & \\
	2009jf & Ib & & \\
	2004gq & Ib & & \\
	\hline
	\end{tabular}
	\label{tab:nebprofile}
\end{table}

An alternative approach to asphericity is considered by \cite{MG2015} with respect to SN 2011fu, whereby they attribute the double peak to a ``blob'' of \ion{O}{} at $\sim4000$ km s$^{-1}$ in the direction of the observer. To investigate this possibility further the nebular profiles of several Type IIb and Ib SNe are plotted in the inset of Figure~\ref{fig:flattop}. Typically they all show some degree of offset with respect to the doublet lines, all in the direction of the observer with velocities of $\sim 2000 - 3000$ km s$^{-1}$, and most have a peak centred on $6300$ \AA . 
Neither approach is wholly convincing because every SNe IIb shows a blueshifted double peaked \ion{[O}{I]} emission (with the exception of SN 2011hs where the \ion{[O}{I]} line is single peaked but shows a high of asymmetry). The probability that all the SNe show motion towards the observer, or are viewed from the same angle, is very small. Yet such a thing would be required for all the nebular \ion{[O}{I]} lines to be blueshifted or double peaked. 
We also conclude that the horns of the \ion{O}{I} 6300, 6363 emission are not due to a doublet component. A similar conclusion was reached by \cite{Modjaz2009} and \cite{Tanaka2009} in relation to SN 2008D, although the latter did attribute the line profile to a toroidal \ion{O}{} distribution contaminated with host \ion{[O}{I]} emission. \cite{Mili2010} investigated the nebular spectra of 5 SE-SNe and concluded that the \ion{[O}{I]} line profile was probably not a consequence of toroidal ejecta or non-spherical geometry \citep[in contrast to][]{Modjaz2008} but instead suggested that the profile could be due to internal scattering or that the far side of the ejecta could be obscured by dust.

It is clear there are a variety of views as to the nature of the double-peak in He-rich SNe, our results, in conjunction with those of \cite{HF1996,Maurer2010} suggest that H$\alpha$ absorption may well be responsible for this feature. The reason for single profile nebular phase \ion{[O}{I]} lines in H-poor He-rich SNe is that there is insufficient densities of H at late times, this could be reflected in the transient nature of the absorption line during the photospheric phase where it disappears before reaching a constant velocity.
The profile of the nebular emission lines for the He-rich SNe in the sample are given in Table~\ref{tab:nebprofile} and it can be seen that the more H-rich SNe show double peaked emission while those with less H can show single-peaked emission.

\subsubsection{Emission}
We now consider if it is possible to see H$\alpha$ emission in the nebular phase. SNe IIb typically show a flat-top emission feature centred on the rest wavelength of H$\mathrm{\alpha}$ which has been attributed to the \ion{[N}{II]} 6548, 6583 doublet \citep{Jerkstrand2015} from N in the He shell, see Figure~\ref{fig:flattop}. However, we take the initial view that this feature could also be due to H$\alpha$, likely as a result of X-rays from ejecta/CSM interaction exciting the unshocked H shell \citep[e.g.,][]{HF1996,Maurer2010,Matheson2000,Maeda2015}. We investigate this possibility by measuring the velocity of the redward edge of the feature, which defines the upper limit of the expansion velocity, although this is sensitive to density. We also consider the edge of the flat top, which represents the inner boundary (hence velocity) of the shell containing the line-forming region. Our results suggest that the apparent ``minimum'' velocity is always lower than the lowest H velocity derived in the photospheric phase. Inspection shows that the maximum and minimum velocities derived for the $\sim6500$ \AA\ region is broadly consistent with the \ion{He}{I} 5876 velocities found in the photospheric phase which would tend to agree with the assessment of \cite{Jerkstrand2015} though some emission component due to H$\alpha$ cannot be ruled out \cite[For a discussion on H$\alpha$ emission in the He shell see][]{Maurer2010}.

\section{He-poor SNe: $\left<N\right>$ in relation to other parameters} \label{sec:7}
$\left<N\right>$ is a measure of line blending, reflecting the density profile of the ejecta above the photosphere which has an effect on $E_\mathrm{k}$/$M_\mathrm{ej}$. However, more information can be extracted from the spectra and from the light curves so there exists a series of properties that can be investigated in relation to $\left<N\right>$ which should help reveal the properties of SE-SNe and the diversity within. From the spectra we can extract line velocities for most SNe (principally those that do not show excessive line blending), or provide estimates on the ejecta velocity, and by using photometric information the temporal and physical characteristics of the LCs can be compared with $\left<N\right>$. 

\subsection{Spectra - $\left<N\right>$ in comparison with line velocities}
\begin{figure*}
	\centering
	\includegraphics[scale=0.7]{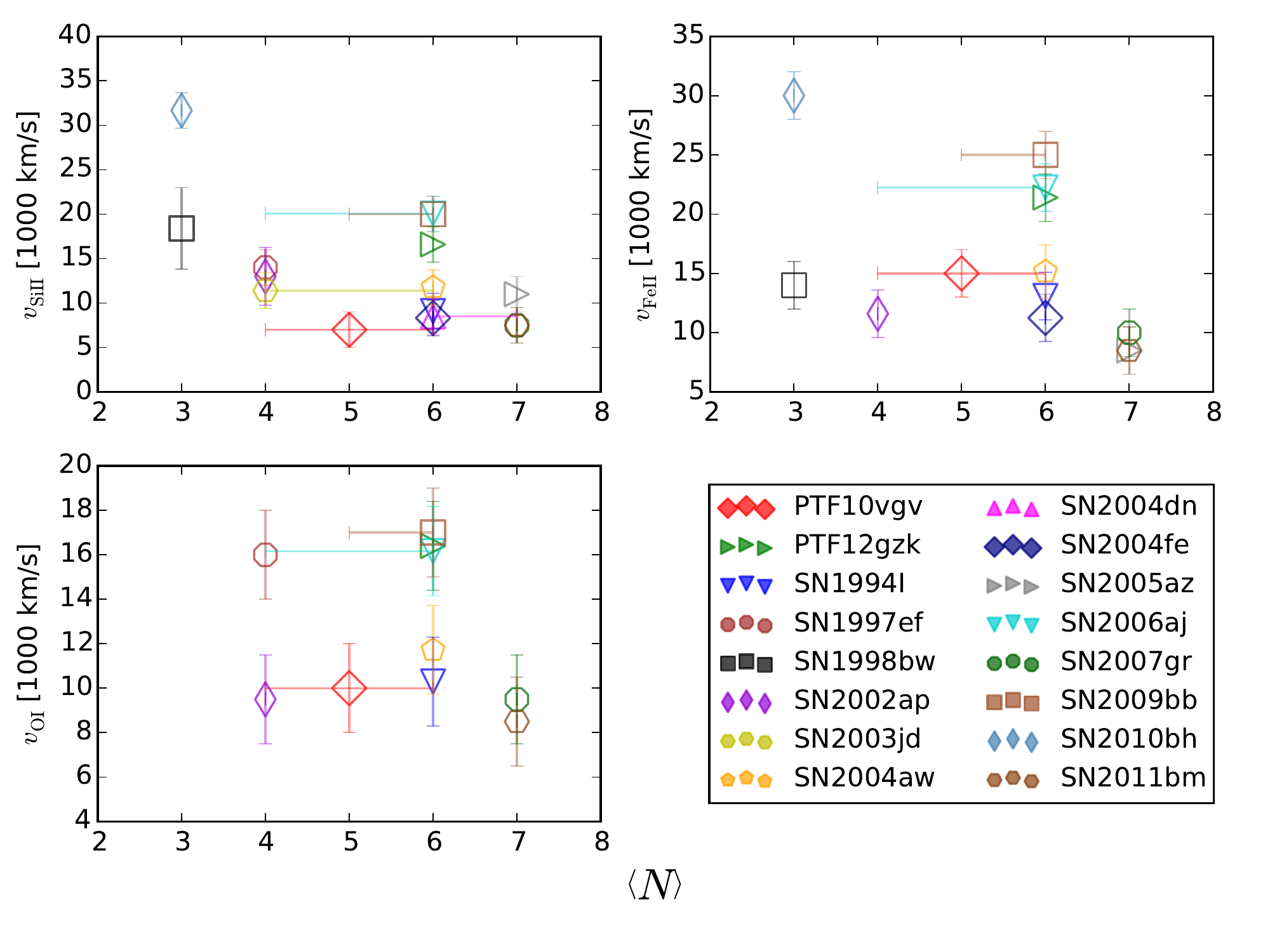}
	\caption{Line velocities around maximum light of Type Ic SNe as a function of $\left<N\right>$. (Top left) The \ion{Si}{II} line velocities show a possible relationship between the velocity of the photosphere and $\left<N\right>$, there is uncertainty with $\left<N\right>$ in some SNe due to contamination from other emission sources, poor S/N, restricted wavelength or temporal coverage. (Top right) The same as before but for \ion{Fe}{II} (note $v_\mathrm{FeII} \approx v_\mathrm{CaII}$), the relationship between line number and the velocity of a line is also present. The number of visible iron lines is related, at least partially, to the iron velocity because this series of lines (\ion{Fe}{II} 5169, 5198, 5235) represents two sources of blending when considering $\left<N\right>$. The positioning of SN 1998bw and SN 2002ap in this plot is interesting because the \ion{Fe}{II} lines in these SNe at this epoch are on the verge of deblending (This is discussed in Section~\ref{sec:unblend}. (Lower left) the \ion{O}{I} lines do not show the same relationship with $\left<N\right>$ and are more scattered, however this is partially because it has not been possible to estimate the \ion{O}{I} velocity when it is blended with the \ion{Ca}{II} NIR triplet.}
	\label{fig:Nvels}
\end{figure*}

Of interest are the line velocities at maximum as a function of $\left<N\right>$, shown in Figure~\ref{fig:Nvels}, of \ion{Si}{II}, \ion{Fe}{II}, and \ion{O}{I}, the first of which is taken as an approximation of the photospheric velocity $v_\mathrm{ph}$. It can be seen that there is a relationship between the velocities of \ion{Si}{II} and \ion{Fe}{II}, although some uncertainty in measurements of $\left<N\right>$ adds to the scatter of the values. This should be expected, because line broadening is usually associated with high energy ejecta which places some material at high velocities. That some SNe show broad lines but their peak velocities settle around the typical value of SNe Ic (SN 1998bw, SN 20002ap) while others retain high velocity lines, either blended (SN 2006aj, SN 2010bh)) or unblended (PTF12gzk) is indicative of the energy distribution in the ejecta. In the former case the energy is primarily ejected into material in the outer layers creating a high velocity region with a flatter density profile. The most energetic SNe tend to be associated with GRBs.

\subsection{$\left<N\right>$ and light curve parameters}
\begin{figure*}
	\centering
	\includegraphics[scale=0.7]{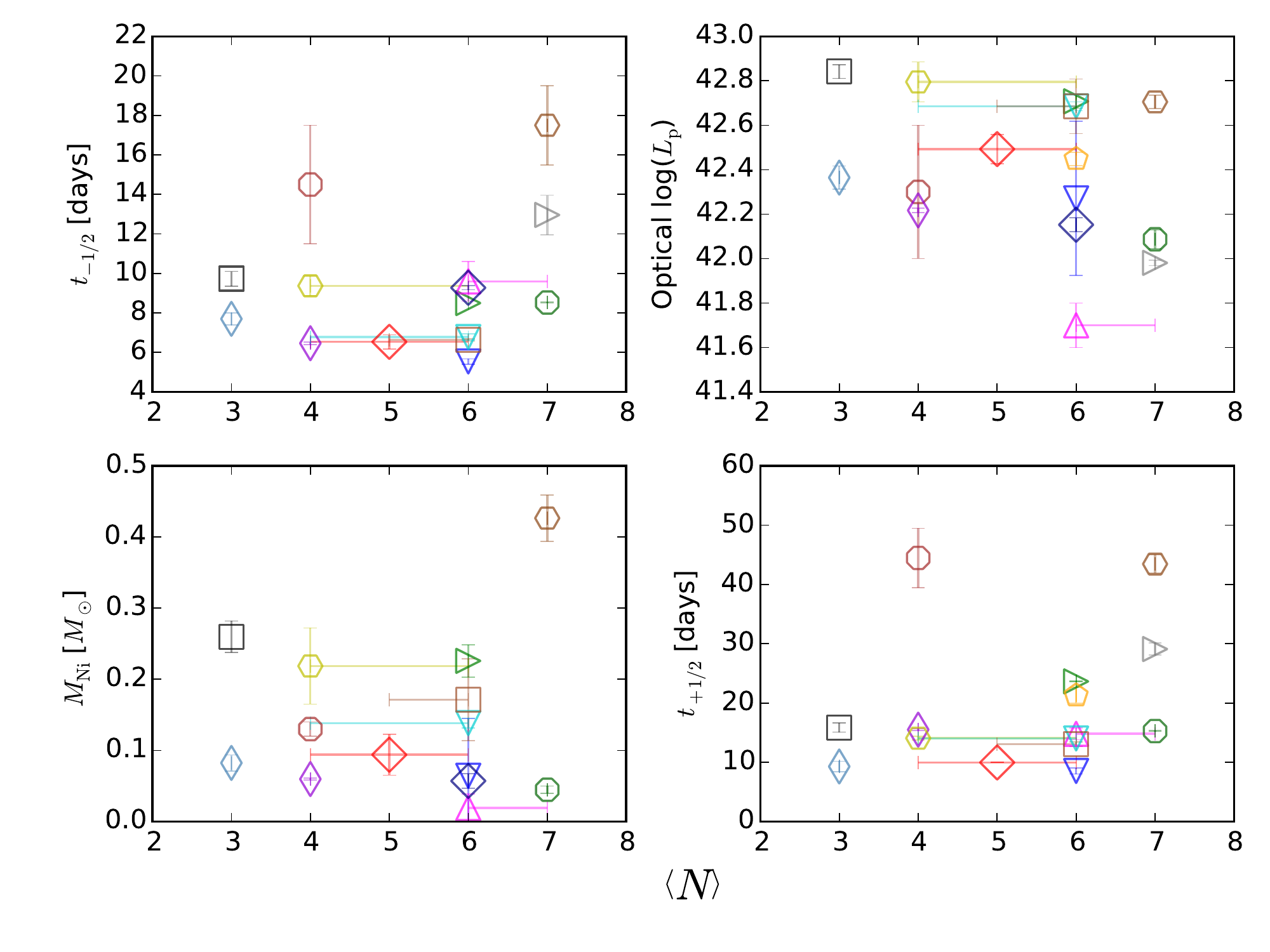}
	\caption{Second parameters when considering $\left<N\right>$. (Top left) $t_{-1/2}$, (top right) optical pseudo-bolometric peak luminosity $L_\mathrm{p}$, (lower left) $M_\mathrm{Ni}$ as derived from the previous value of $L_\mathrm{p}$, (lower right) \tdecay. \lp\ and \mni\ include host-galaxy extinction when known but some are lower limits, SN 2004dn in particular appears heavily extinguished. None of the parameters individually show a correlation with $\left<N\right>$ but they are all related to each other through the photon diffusion and ejecta expansion times cales \citep{Arnett1982}. SN labels as the same as in Figure~\ref{fig:Nvels}}
	\label{fig:Neverything}
\end{figure*}

Having considered how various line velocities relate to $\left<N\right>$ we can investigate the relationship between $\left<N\right>$ and properties of the light curve: $t_{-1/2}$, $L_\mathrm{p}$, $M_\mathrm{Ni}$, and \tdecay\ which is plotted in Figure~\ref{fig:Neverything}. The values for these three properties are taken from \cite{Prentice2016} and their derivation described therein. The optical ($4000 - 10000$ \AA ) pseudo-bolometric $L_\mathrm{p}$ is used here in order to maximise the available SNe.

It appears that $\left<N\right>$ is weakly correlated with \lp, however this is driven by the GRB-SNe, which are exclusively Ic-3 and significantly more luminous on average than other He-poor SNe, and the choice of SNe used here. In the absence of the GRB-SNe there is no relation between \lp\ and $\left<N\right>$, nor is there a correlation with any of the other parameters investigated.

\section{Calculating \ek} \label{sec:ek}
We note that deriving an accurate value for \ek\ is not trivial. For example, It is common for $E_\mathrm{k}$ to be calculated using the well known equation from \citet{Arnett1982}

\begin{equation}
E_\mathrm{k}=3/10M_\mathrm{ej}v_\mathrm{sc}^2
\label{eqn:arnettEK}
\end{equation}

Where $v_\mathrm{sc}$ is a scale velocity. Equation~\ref{eqn:arnettEK} assumes a homologous (e.g, $R(v,t)=vt$), spherical expansion. In this model the ejecta density $\rho$ is constant whereas models typically have the outer density profile take the form of $\rho(r)\propto r^{-n}$ where $n$ is usually $\sim7$ \citep{Mazzali2000}. 
Additionally, this model cannot account for changes in the density distribution so for SNe where more mass is placed at higher velocities (those that show long lived broad lines) $E_\mathrm{k}$ suffers from large uncertainties. 
The final point here is that $v_\mathrm{sc}$ is commonly taken to be the photospheric velocity at maximum, a quantity that is not measurable from the spectra but is derived from modelling. However, the derivation of Equation~\ref{eqn:arnettEK} leads to $v_\mathrm{sc}=v_\mathrm{max}$. Thus, as $E_\mathrm{k}$ is related to $v_\mathrm{sc}^{2}$ then the kinetic energy derived is very sensitive to the choice of $v_\mathrm{sc}^{2}$, which leads to large uncertainties. Because analytical methods cannot account for varying density and an uncertain characteristic velocity \citep{Mazzali2013}, a proper treatment of $E_\mathrm{k}$ requires spectral modelling at early times \citep[e.g.,][]{Mazzali2000b} or hydrodynamical modelling of the explosion \citep[for example,][]{Iwamoto1998}. 

\section{Tables}
\begin{table*}
	\caption{Database }
	\begin{tabular}{cc c c c}
	
SN & $z$ & Original classification & Reclassification & References \\
SN1993J & -0.0001 & IIb & IIb &(1) (2)  \\
SN1994I & 0.0015 & Ic & Ic-6(11/9) & (3) (4)  \\
SN1996cb & 0.002 & IIb & IIb(I) &(3) (5)  \\
SN1997ef & 0.012 & Ic-BL & Ic-4(13/45) & (3) (5) (6) (7)  \\
SN1998bw & 0.0085 & GRB-SN & Ic-3(15/16) & (8)  \\
SN1999dn & 0.0093 & Ib & Ib(II) &(9) (10) (5) (11)  \\
SN1999ex & 0.011 & Ib & Ib(II) & (12)  \\
SN2002ap & 0.0022 & Ic-BL & Ic4-9/6 & (13) (14) (3) (15) (16)  \\
SN2003bg & 0.0046 & IIb & IIb & (17) (18)  \\
SN2003jd & 0.019 & Ic-BL & Ic-4(13/14) & (3)  \\
SN2004aw & 0.016 & Ic & Ic-6(11/21) & (3) (19)  \\
SN2004dn & 0.013 & Ic & Ic-6(9/15) &(3) (20)  \\
SN2004fe & 0.018 & Ic &Ic-6(8/u) &(20) (3)  \\
SN2004gq & 0.0065 & Ib & Ib & (3)  \\
SN2005az & 0.0085 & Ic &Ic-7(11/29)& (3)  \\
SN2005bf & 0.019 & Ib & Ib(II) &(21) (3)  \\
SN2005hg & 0.021 & Ib & Ib & (3)  \\
SN2006T & 0.008 & IIb & IIb& (3)  \\
SN2006aj & 0.033 & XRF-SN & Ic-6(21/14) & (22) (23) (3)  \\
SN2006el & 0.017 & IIb & IIb(I) & (3)  \\
SN2006ep & 0.015 & Ib & Ib(II)& (3)  \\
SN2006lc & 0.016 & Ib & Ib & (3)  \\
SN2007Y & 0.0046 & Ib & Ib(II) & (24)  \\
SN2007gr & 0.0017 & Ic & Ic-7(7/15) & (25) (3)  \\
SN2007kj & 0.018 & Ib & Ib(II) & (3)  \\
SN2007uy & 0.0065 & Ib & Ib(II) & (3)  \\
SN2008D & 0.0065 & Ib & Ib(II) & (26) (27) (28)  \\
SN2008ax & 0.0019 & IIb & IIb(I) & (3) (29) (30) (31)  \\
SN2008bo & 0.005 & IIb & IIb & (3)  \\
SN2009bb & 0.0099 & Ic-BL & Ic-6(20/13) & (32)  \\
SN2009er & 0.035 & Ib & Ib &  (3)  \\
SN2009iz & 0.014 & Ib & Ib & (3)  \\
SN2009jf & 0.0079 & Ib & Ib & (33) (3)  \\
PTF10vgv & 0.015 & Ic & Ic-5(7/10) & (34)  \\
SN2010ah & 0.050 & Ic-BL & Ic-3(18/17) &(35)  \\
SN2010as & 0.007 & IIb & IIb(I) & (36)  \\
SN2010bh & 0.059 & GRB-SN & Ic-3(32/9) & (37)  \\
SN2011bm & 0.022 & Ic & Ic-7(6/43) & (38)  \\
SN2011dh & 0.002 & IIb & IIb & (39) (40)  \\
SN2011ei & 0.0093 & IIb & IIb & (41)  \\
SN2011fu & 0.0185 & IIb & IIb & (42) (43)  \\
SN2011hs & 0.0057 & IIb & IIb & (44)  \\
PTF12gzk & 0.0137 & Ic & Ic-6(17/24) & (45)  \\
SN2012ej & 0.009 & Ic &  Ic-7(7/20) & (48)\\
iPTF13bvn & 0.0045 & Ib & Ib & (46) (47)  \\
SN2016P & 0.0146 & Ic-BL & Ic-6(u/14) & (48) \\
SN2016coi & 0.0036 & Ic-BL & Ic-4(14/21) &(48) \\
SN2016iae & 0.004 & Ic &Ic-7(9/14)& (48) \\
\multicolumn{5}{p{\textwidth}}{References: (1) \citep{Matheson2000}, (2) \citep{Barbon1995}, (3) \citep{Modjaz2014}, (4) \citep{Filippenko1995}, (5) \citep{Matheson2001}, (6) \citep{Iwamoto2000}, (7) \citep{Mazzali2000}, (8) \citep{Patat2001}, (9) \citep{Deng2000}, (10) \citep{Benetti2011}, (11) \citep{Taubenberger2009}, (12) \citep{Hamuy2002}, (13) \citep{GalYam2002}, (14) \citep{Chornock2013}, (15) \citep{Mazzali2002}, (16) \citep{Foley2003}, (17) \citep{Hamuy2009}, (18) \citep{Mazzali2009}, (19) \citep{Taubenberger2006}, (20) \citep{Harutyunyan2008}, (21) \citep{Folatelli2006}, (22) \citep{Pian2006}, (23) \citep{Sonbas2008}, (24) \citep{Stritzinger2009}, (25) \citep{Valenti2008}, (26) \citep{Modjaz2009}, (27) \citep{Malesani2009}, (28) \citep{Mazzali2008}, (29) \citep{Taubenberger2011}, (30) \citep{Pastorello2008}, (31) \citep{Mili2010}, (32) \citep{Pignata2011}, (33) \citep{Valenti2011}, (34) \citep{Corsi2012}, (35) \citep{Corsi2011}, (36) \citep{Folatelli2014}, (37) \citep{Bufano2012}, (38) \citep{Valenti2012}, (39) \citep{Arcavi2011}, (40) \citep{Ergon2014}, (41) \citep{Mili2013}, (42) \citep{MG2015}, (43) \citep{Kumar2013}, (44) \citep{Bufano2014}, (45) \citep{Benami2012}, (46) \citep{Cao2013}, (47) \citep{Srivastav2014}, (48) (Prentice et al., in preparation)}\\
	\hline
	\end{tabular}
	\label{tab:database}
\end{table*}

\bsp	
\label{lastpage}
\end{document}